\documentclass[a4paper,usenatbib]{mnras}
\usepackage{graphicx}
\usepackage[fleqn]{amsmath}
\usepackage[T1]{fontenc}
\usepackage{aecompl}
\usepackage{times}
\usepackage{ae,aecompl}
\usepackage{amssymb,amsfonts,hyperref,xcolor}
\usepackage{bm}

\usepackage[encapsulated]{CJK}
\usepackage{ucs}
\usepackage[utf8x]{inputenc}
\newcommand{\cntext}[1]{\begin{CJK}{UTF8}{bkai}#1\ignorespacesafterend\end{CJK}}  

\newcommand\be{\begin{equation}}
\newcommand\en{\end{equation}}

\renewcommand{\vec}{\bm}
\newcommand{\unitvec}[1]{\hat{\vec{e}}_{#1}}
\newcommand{\rhop}{\rho_\mathrm{p}}
\newcommand{\mass}{m_\mathrm{p}}
\newcommand{\AR}{A_\mathrm{R}}
\newcommand{\AI}{A_\mathrm{I}}
\newcommand{\kp}{\vec{k}_+}
\newcommand{\km}{\vec{k}_-}
\newcommand{\rj}{\vec{r}_j}
\newcommand{\xj}{\vec{\xi}_j}

\newcommand{\pc}{\textsc{Pencil Code}}

\newcommand{\tder}[2]{\frac{\mathrm{d}#1}{\mathrm{d}#2}}
\newcommand{\Ts}{T_\mathrm{s}}
\newcommand{\Tsi}[1]{T_{\mathrm{s},#1}}
\newcommand{\Tsmin}{T_{\mathrm{s,min}}}
\newcommand{\Tsmax}{T_{\mathrm{s,max}}}
\newcommand{\Nsp}{N_\mathrm{sp}}
\newcommand{\OmegaK}{\Omega_\mathrm{K}}
\newcommand{\vK}{v_\mathrm{K}}
\newcommand{\cs}{c_\mathrm{s}}

\newcommand{\phn}{\phantom{0}}

\title[Multi-species SI]{Streaming Instability with Multiple Dust Species: I.~Favourable Conditions for the Linear Growth}

\author[Z.~Zhu \& C.-C.~Yang]{%
Zhaohuan Zhu \cntext{(朱照寰)}$^{1}$\thanks{E-mail: zhaohuan.zhu@unlv.edu} and Chao-Chin Yang \cntext{(楊朝欽)}$^{1}$\\
$^{1}$Department of Physics and Astronomy, University of Nevada, Las Vegas, 4505 S.~Maryland Parkway, Box~454002,\\Las Vegas, NV~89154-4002, USA}

\date{In original form \today}

\begin{document}
\label{firstpage}
\pagerange{\pageref{firstpage}--\pageref{lastpage}} \pubyear{2019}
\maketitle

\begin{abstract}
Recent study suggests that the streaming instability, one of the leading mechanisms for driving the formation of planetesimals, may not be as efficient as previously thought.
Under some disc conditions, the growth timescale of the instability can be longer than the disc lifetime when multiple dust species are considered.
To further explore this finding, we use both linear analysis and direct numerical simulations with gas fluid and dust particles to mutually validate and study the unstable modes of the instability in more detail.
We extend the previously studied parameter space by one order of magnitude in both the range of the dust-size distribution $[\Tsmin, \Tsmax]$ and the total solid-to-gas mass ratio $\varepsilon$ and introduce a third dimension with the slope $q$ of the size distribution.
 {We find that the fast-growth regime and the slow-growth regime are distinctly separated in the $\varepsilon$--$\Tsmax$ space, while this boundary is not appreciably sensitive to $q$ or $\Tsmin$. With a wide range of dust sizes present in the disc (e.g.\ $\Tsmin\lesssim10^{-3}$), the growth rate in the slow-growth regime decreases as more dust species are considered.
With a narrow range of dust sizes (e.g. $\Tsmax/\Tsmin=5$), on the other hand, the growth rate in most of the $\varepsilon$--$\Tsmax$ space is converged with increasing dust species, but the fast and the slow growth regimes remain clearly separated.}
Moreover, it is not necessary that the largest dust species dominate the growth of the unstable modes, and the smaller dust species can affect the growth rate in a complicated way.
In any case, we find that the fast-growth regime is bounded by $\varepsilon\gtrsim1$ \emph{or} $\Tsmax\gtrsim1$, which may represent the favourable conditions for planetesimal formation.
\end{abstract}

\begin{keywords}
hydrodynamics -- planets and satellites: formation -- protoplanetary discs -- instabilities 
\end{keywords}

\section{Introduction}
One major obstacle for planet formation under the core accretion scenario is how solids can grow from cm/m-sized objects to km-sized planetesimals within a 1--10\,Myr disc lifetime \citep{Williams2011,Ansdell2016}.
Dust can grow from the ISM $\mu$m sizes to mm/cm sizes via electrostatic forces at collisions.
Objects of km sizes and above can bind themselves and accrete materials via their own gravity \citep{BA99,LS09,BB17,SD20}.
However, it is difficult for solids to grow from pebble sizes to km sizes \citep{Johansen2014}.
Collisions between dust at these sizes often lead to bouncing or fragmentation instead of growth \citep{Blum2008,ZO10}.
Furthermore, mm/cm-sized pebbles in the outer disc or meter-sized boulders in the inner disc radially drift inward problematically fast in protoplanetary discs.
In a smooth disc, these solids should drift to the central star within $\sim$10$^4$ years, which is known as the radial-drift barrier for planetesimal formation \citep{AHN76,Weidenschilling1977,BDH08,BFJ16}.

Several mechanisms have been proposed to overcome these barriers to planetesimal formation, including dust traps at a local pressure maximum \citep{Whipple1972, Johansen2009,BS14} or within vortices \citep{Barge1995,Lyra2009}, growth of porous icy dust aggregates \citep{Kataoka2013}, secular gravitational instability \citep{Youdin2011,MKI12,Takahashi2014}, and the streaming instability \citep{YoudinGoodman2005,YJ07,Johansen2007,Lin2017}.
Among these, the streaming instability is one promising mechanism to concentrate dust particles of a wide range of sizes (\citealt{CJD15,Yang2017}; R.~Li et al., in preparation).
Unlike other mechanisms, which passively rely upon the underlying disc structures, dust particles actively participate in the dust-gas dynamics and spontaneously concentrate themselves via the reaction of the aerodynamic drag back onto the gas \citep{YJ14,LYS18}.

However, \cite{Krapp2019} found that the linear growth of the streaming instability can be significantly different
when considering a range of dust sizes instead of a single dust species as in previous studies.
For some dust-size distributions and total solid-to-gas mass ratios, the growth rate of the instability decreases with increasing number of discrete species representing the distribution.
It appears that for these cases, the growth timescale can be much longer than the typical disc lifetime for large number of dust species.
This implies that taking the limit of a continuous dust-size distribution, the instability may not operate at all in protoplanetary discs.
On the other hand, previous stratified numerical simulations \citep{BS10,SYJ18} have indicated that the streaming instability with multiple dust sizes appears to operate in these stratified disks and can even drive strong concentration of dust particles.
Furthermore, the effects of background turbulence have not been considered in this scenario yet \citep{YMJ18,CL20,UEC20}, and it appears that turbulence may play an important role in determining the efficiency and scale of planetesimal formation \citep{GM20,GS20,KS20}.
Therefore, further studies of the multi-species streaming instability seem warranted.

In this work, we augment the study by \cite{Krapp2019} as follows.
We validate the linearly unstable modes of the instability by using both linear analysis and direct numerical simulations (Sec.~\ref{S:methods}).
We expand the parameter space by one order of magnitude in both the range of dust sizes and the total solid-to-gas ratio, and moreover, we explore the effects of the slope of the dust-size distributions (Sec.~\ref{S:tworeg}).
 {Then, we discuss the properties of the most unstable mode (Sec.~\ref{S:wavenumber}), the contributions from different dust species (Sec.~\ref{S:interaction}), long wavelength modes in a special region of the parameter space (Sec.~\ref{SS:long}), and the instability with a narrow dust size distribution (Sec.~\ref{S:para}).}
Finally, we briefly summarize our results (Sec.~\ref{S:summary}).

\section{Methods} \label{S:methods}
\subsection{Linear analysis\label{SS:la}}
We carry out the linear analysis for the multi-species streaming instability using the linearised perturbation equations presented in Appendix~E of \cite{Benitez2019}, which assumes no vertical stratification and no background turbulence.
The underlying dust-size distribution is assumed to follow  {a power law $dn(s)/ds\propto s^{q}$} \citep{Mathis1977} from $s_\mathrm{min}$
to $s_\mathrm{max}$, where $s$ is the size of the dust particle.
Since dust particles under the conditions of a typical protoplanetary disc are mostly in the Epstein regime  {\citep[e.g.,][]{Johansen2014}}, we assume that the dimensionless stopping time ($\Ts$)  {of the particles} is proportional to  {their} size.

Under these assumptions, we consider dust with sizes from $\Tsmin$ to $\Tsmax$ (i.e., $\Ts \in [\Tsmin, \Tsmax]$) and a total dust-to-gas mass ratio of $\varepsilon$ in a disc.
To discretise the continuous distribution into $\Nsp$ dust species, we divide the dust distribution into $\Nsp$ bins uniformly in the $\log\Ts$ space.
The dust particles in each bin have identical stopping time $\Tsi{i}$, being at the linear center of the bin.
The dust-to-gas mass ratio for each bin is
\begin{equation} \label{E:epsj}
    \varepsilon_{i} = \frac{\Tsi{i,\textrm{u}}^{4+q}-\Tsi{i,\textrm{l}}^{4+q}}{\Tsmax^{4+q}-\Tsmin^{4+q}}\varepsilon,
\end{equation}
where $\Tsi{i,\textrm{u}}$ and $\Tsi{i,\textrm{l}}$ are the upper and lower dust size limits for each dust bin.
 {With Equation~\eqref{E:epsj}, $\varepsilon=\Sigma_i\varepsilon_i$, and hence when} we increase the number of species $\Nsp$ with the same distribution, the total mass of the dust remains constant.

Under the local-shearing-box approximation, \cite{Benitez2019} derived the equations describing the gas and $\Nsp$ dust species in the radial-vertical ($xz$) plane assuming axisymmetry.
The linearised equations have been written in terms of the perturbed variables $\delta \hat{f}(k_x,k_z) e^{i(k_x x+k_z z)-\omega t}$ in the Fourier space.
The wave number ($k_x$, $k_z$) and the complex eigenvalue $\omega(k_x,k_z)$ can be expressed in the dimensionless form of $K_i = k_i\eta R_0$ with $i = x, z$ and $\tilde{\omega}=\omega/\OmegaK(R_0)$, where $\eta$ is related to the radial pressure gradient at the orbital radius $R = R_0$ by
\begin{equation}
\eta\equiv\frac{h_0^2}{2}\left.\tder{\log P}{\log R}\right|_{R_0},
\end{equation}
$h_0 \equiv H_0 / R_0$ with $H_0$ being the disc scale height at $R_0$, and $\OmegaK(R)$ is the Keplerian angular frequency \citep{NSH86}.
We choose $\eta \vK / \cs = 0.05$ as our fiducial radial pressure gradient, where $\vK = R\OmegaK$ and $\cs$ are the Keplerian velocity and the local speed of sound at $R_0$, respectively.
We have also tried $\eta \vK / \cs = 0.1$ for the outer disc, but the results are almost identical to those with the fiducial gradient.
The eigenvalues $\omega$ of the linearised equations are computed using Python NumPy function {\ttfamily linalg.eigvals}, and the growth rate of the multi-species streaming instability is hence $\sigma = -\mathrm{Re}(\omega)$.
All growth rates $\sigma$  {presented in this article is normalised by} $\Omega_K$.
 {We have compared our growth rates to all linear calculations considered in \cite{YJ07} and \cite{Benitez2019}, and the agreement on the growth rate and each eigenmode is more than six significant digits.}

We consider three different power-law indices $q = -3.5, -2.5$, and $-1.5$ for the dust-size distribution, motivated by protoplanetary disc observations and dust coagulation/fragmentation calculations (e.g., \citealt{Perez2015}, \citealt{Birnstiel2012}).
The smallest dust particles are fixed at $\Tsmin = 10^{-4}$ or $10^{-3}$  {for most cases, while we change $\Tsmin$ to be some fraction of $\Tsmax$ in Sec.~\ref{S:para}.}
Given the maximum dust stopping time $\Tsmax$ and the total solid-to-gas mass ratio $\varepsilon$, we calculate the maximum growth rate from all the eigenmodes for each $K_x$ and $K_z$.
We then search for the maximum growth rate among the Fourier space $(K_x, K_z)$.
This space is infinite, so we need to confine our search in practice.
\cite{YoudinGoodman2005} have suggested that $K_x \sim K_z \sim 1 / \Ts$ is roughly where the fastest growing mode is for the single-species streaming instability.
Therefore, centred around $K_x = K_z = 1 / \Tsmax$, we uniformly choose 54 values of $\log_{10}K_x$ in $[A-1.5, A+2.5]$ and 54 values of $\log_{10}K_z$ in $[A-3, A+3]$, where $A \equiv -\log_{10}\Tsmax$.
By inspection, we find that this domain size generally captures the absolute maximum of the growth rate.
 {We notice though that the fastest growing mode for some cases has appreciably small $K_x$ and $K_z$ with respect to $1/\Tsmax$ (Sec.~\ref{S:wavenumber}).
Thus, we consider a different domain size for these cases with $\log_{10}K_x$ in $[A-3, A+2]$ and $\log_{10}K_z$ in $[A-5, A+2]$ in Sec.~\ref{SS:long}.}
Finally, we define the maximum growth rate $\sigma$ for any given $\Tsmax$ and $\varepsilon$ as the maximum among the 54$\times$54 computed growth rates.
 {Due to the high computational cost of solving the eigenvalue problem for up to 1024 dust species, we decide to limit to 54$\times$54 values for the large $K_x$-$K_z$ space.
Furthermore, \cite{Krapp2019} noticed some fine structures in the $K_x$-$K_z$ growth rate map with multiple dust species in the disc.
Although our wave number resolution and the potential fine structures have little effect on the maximum growth rate, they can affect the exact values of $K_x$ and $K_z$ for the fastest growing mode, which may lead to minor non-smooth profiles in some of our figures.}

For each given dust distribution, we systematically increase the number of dust species $\Nsp$ from 2 to 1024 and use the aforementioned method to find the maximum growth rate $\sigma$ for each $\Nsp$.
However, the linear algebra becomes quite computationally intensive when $\Nsp$ is more than 1024.
To estimate the maximum growth rate with more dust species (e.g., $\Nsp = 2048$ and $\Nsp = 4096$), therefore, we simply use the $K_{x}$ and $K_{z}$ of the fastest growing mode found with 1024 dust species and only calculate the growth rate for this combination of $K_{x}$ and $K_{z}$ for 2048 and 4096 dust species, leading to our $\sigma_{2048}$ and $\sigma_{4096}$, respectively.
Our experience with $\Nsp \leq 1024$ is that, in most cases, increasing $\Nsp$ does not change the wave number of the fastest growing mode, so our approach should give a good approximation of the maximum growth rate with 2048 and 4096 dust species.  {However, the fastest growing mode in some cases do change location in the $K_x$-$K_z$ space when we increase the number of dust species (Sec.~\ref{S:wavenumber} and~\ref{S:interaction}).
Thus, we focus on $\sigma_{1024}$ in this work and cautiously present results for $\sigma_{2048}$ and $\sigma_{4096}$ where they are available.}

\subsection{Numerical validation\label{SS:nv}}
In addition to linear analysis, we use a simulation code to reproduce a linear mode for several cases, which serves for two purposes.
First, the results from our linear analysis and the code are mutually validated, and in this process we demonstrate the resolution requirement for simulating the streaming instability with multiple dust species.
Second and perhaps more importantly, by using a code equipped with Lagrangian dust particles, we can gauge the validity of the multi-fluid approximation for dust particles used in the linear analysis, especially for relatively large ones ($\Ts \sim 1$, \citealt{GBL04}) which also happen to have the highest growth rates.

For these purposes, we use the \pc{}\footnote{The \pc{} and its documentation is publicly available at \url{http://pencil-code.nordita.org/}.} \citep{BD02}.
It employs sixth-order finite differences to approximate any spatial derivatives and integrates the system of (magneto-)hydro-dynamical equations in time using third-order Runge--Kutta method.
To stabilize the scheme, we use sixth-order hyper-diffusion with fixed Reynolds number \citep{YK12}.
The dust component is modelled as Lagrangian super-particles, and their trajectories are integrated in tandem with the Runge--Kutta steps.
The drag interaction between each dust particle and its surrounding gas is achieved via the standard particle-mesh method \citep{YJ07}.

For each dust distribution $[\Tsmin, \Tsmax]$ and total solid-to-gas density ratio $\varepsilon$ we explore, we select a mode with a wave number $K_x = K_z$ that is close to the fastest growing mode.
We normalize the eigenvector such that the amplitude of the perturbation in \emph{total} particle density relative to the equilibrium density is about $10^{-6}$.
We use a square computational domain that accommodates one wavelength per dimension.
While it is straightforward to seed the mode in the gas component with Eulerian formulation, it is not trivial to do so in the dust component with Lagrangian super-particles.
We describe in detail in Appendix~\ref{S:seedpar} the algorithm of how to position the particles to focus the perturbations onto a desired mode.
The velocities of the particles are then assigned according to their positions.
For all cases, we allocate four particles per species per cell.%
\footnote{%
    We find that with one particle per species per cell, grid noise of small amplitudes appear at high resolutions.
    Using four particles per cell effectively eliminates this noise.}
With that, we conduct each simulation up to a time on the order of one $e$-folding time, and measure the growth rate of the mode in each field and compare it with the theoretical growth rate obtained from the linear analysis.

\section{Two regimes of the instability} \label{S:tworeg}

We explore the 3D parameter space spanned by the maximum dust stopping time $\Tsmax$, the total solid-to-gas mass ratio $\varepsilon$, and the power-law index $q$ for the dust-size distribution. For our main set of calculations, we fix
 the minimum dust size with $\Tsmin = 10^{-4}$, select 21 different $\log_{10}\Tsmax$ values from $-3$ to 1 and 16 different $\log_{10}\varepsilon$ values from $-2$ to 1.
Therefore, both our $\Tsmax$ and our $\varepsilon$ are one order of magnitude larger than those investigated by \cite{Krapp2019}.
Furthermore, we expand their parameter study with a third dimension by considering three different $q = -3.5, -2.5$, and $-1.5$.
We describe our findings in this section.

\subsection{Characteristic conditions\label{SS:cc}}

\begin{figure*}
\includegraphics[trim=0mm 1mm 0mm 0mm, clip, width=6.6in]{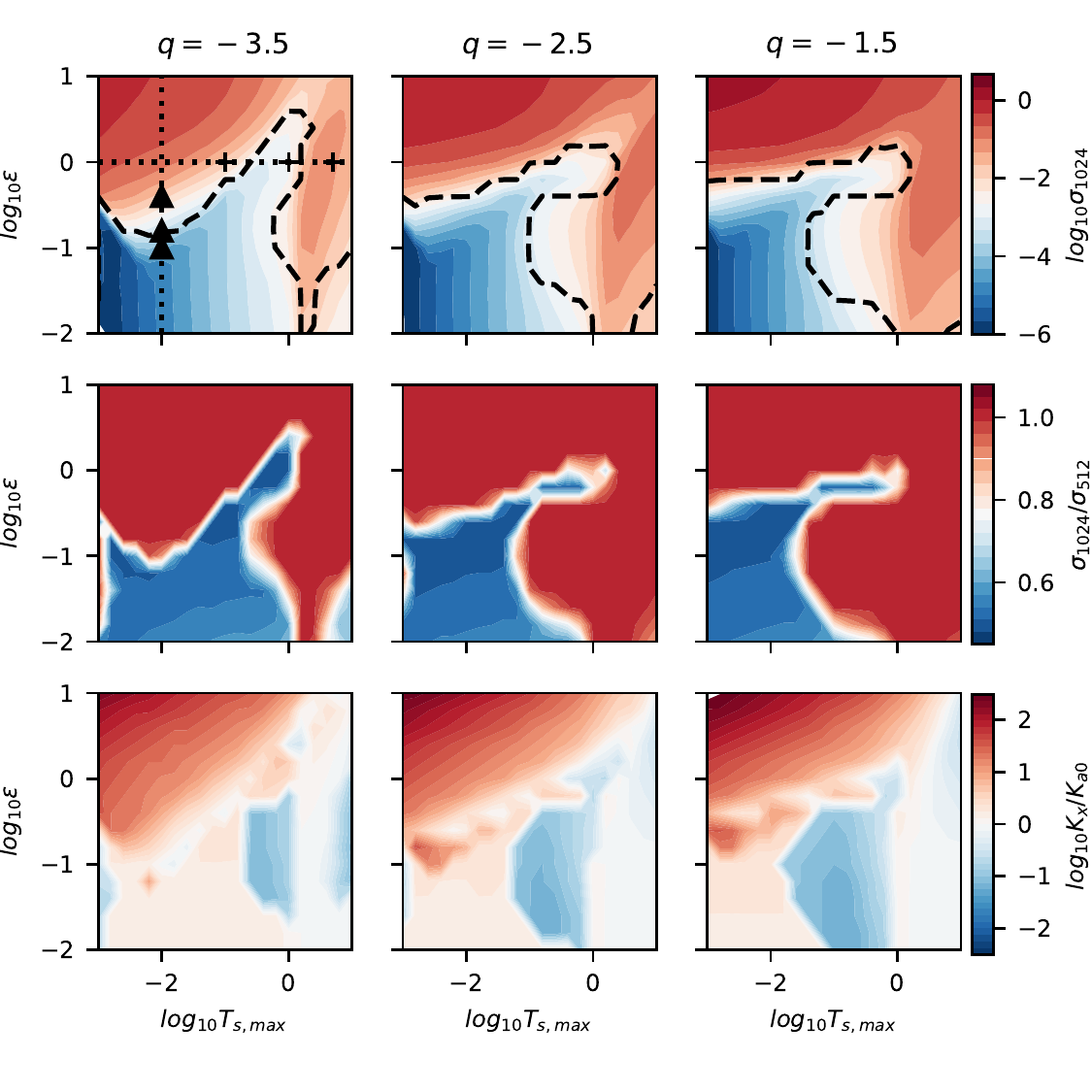}
\caption{%
    Maximum growth rate $\sigma$ (in $\OmegaK$) of the multi-species streaming instability for discs with different maximum dust size at $\Tsmax$, total solid-to-gas mass ratio $\varepsilon$, and power-law index $q$ of the dust-size distribution. $\Tsmin$ is fixed at $10^{-4}$.
    The higher $q$ (less negative), the more top-heavy the distribution is.
    The upper panels show the rate with 1024 dust species $\sigma_{1024}$.
    The middle panels show the ratio between the rates with 1024 and 512 dust species $\sigma_{1024} / \sigma_{512}$.
    The dashed lines in the upper panels denote $\sigma_{1024} / \sigma_{512} = 0.99$.
    The dotted lines are the 1D cuts analyzed in Sec.~\ref{SS:transition}.
    The triangles and the crosses mark the parameter sets studied in Sec.~\ref{S:wavenumber}.
    The bottom panels show the radial component of the dimensionless wave number $K_{x}$ of the fastest growing mode with 1024 dust species. $K_x$ is normalized by $K_{a0}$ (Equation \ref{eq:Kx}).}
\label{fig:multi1}
\end{figure*}

The top-left panel in Fig.~\ref{fig:multi1} shows the maximum growth rate $\sigma_{1024}$ for 1024 dust species following the dust-size distribution with the power-law index $q = -3.5$.
Restricting our attention to the parameter space of $\Tsmax \le 1$ and $\varepsilon \le 1$, we have reproduced the growth rate found by \cite{Krapp2019} (see their Fig.~5).
The growth rate generally increases with increasing $\varepsilon$.
Two distinct regimes appear separated by roughly $\varepsilon \sim 0.3$, above which the growth rate of the instability is high ($\sigma \gtrsim 0.01\OmegaK$) and below which it is appreciably lower ($\sigma \ll 0.01\OmegaK$).
As we expand the parameter space up to $\varepsilon = 10$, we find that the trend continues, as $\sigma$ increases with increasing $\varepsilon$.
Moreover, we find that $\sigma$ also increases with decreasing $\Tsmax$ above the transition.
The maximum growth timescale is shorter than the orbital time near our smallest $\Tsmax \simeq 10^{-3}$ and our largest $\varepsilon \simeq 10$.

More interestingly, as we increase the maximum dust size up to $\Tsmax = 10$, we find another transition zone where the maximum growth rate changes from low to high.
As shown by the top-left panel in Fig.~\ref{fig:multi1}, this transition lies at roughly $\Tsmax \sim 1$.
Therefore, it appears that high growth rate occurs when \emph{either} the maximum size $\Tsmax$ \emph{or} the total solid-to-gas mass ratio $\varepsilon$ is high, while low growth rate occurs when \emph{both} $\Tsmax$ \emph{and} $\varepsilon$ are low.

We further investigate the effects of changing the slope of the dust-size distribution.
The top panels in Fig.~\ref{fig:multi1} compares the maximum growth rate $\sigma_{1024}$ for different power-law index $q$, with flatter size distribution (more top-heavy) toward the right (see Eq.~\ref{E:epsj}).
As shown by the panels, the general trend found above remains the same, while there are only slight changes of the transition zone separating fast and slow growth of the instability.
With a more top-heavy dust distribution, the transition at $\varepsilon \sim 1$ becomes less sensitive to $\Tsmax$, and there is a slightly larger parameter space at $\Tsmax \sim 1$ for high growth rates. On the other hand, the most unstable modes for cases at $0.1\lesssim \Tsmax\lesssim 1$ have small $K_z$ which may not fit into the disc thickness, which  {we discuss in more detail} in Sec.~\ref{S:wavenumber}.

\begin{figure*}
\includegraphics[trim=0mm 1mm 0mm 0mm, clip, width=6.in]{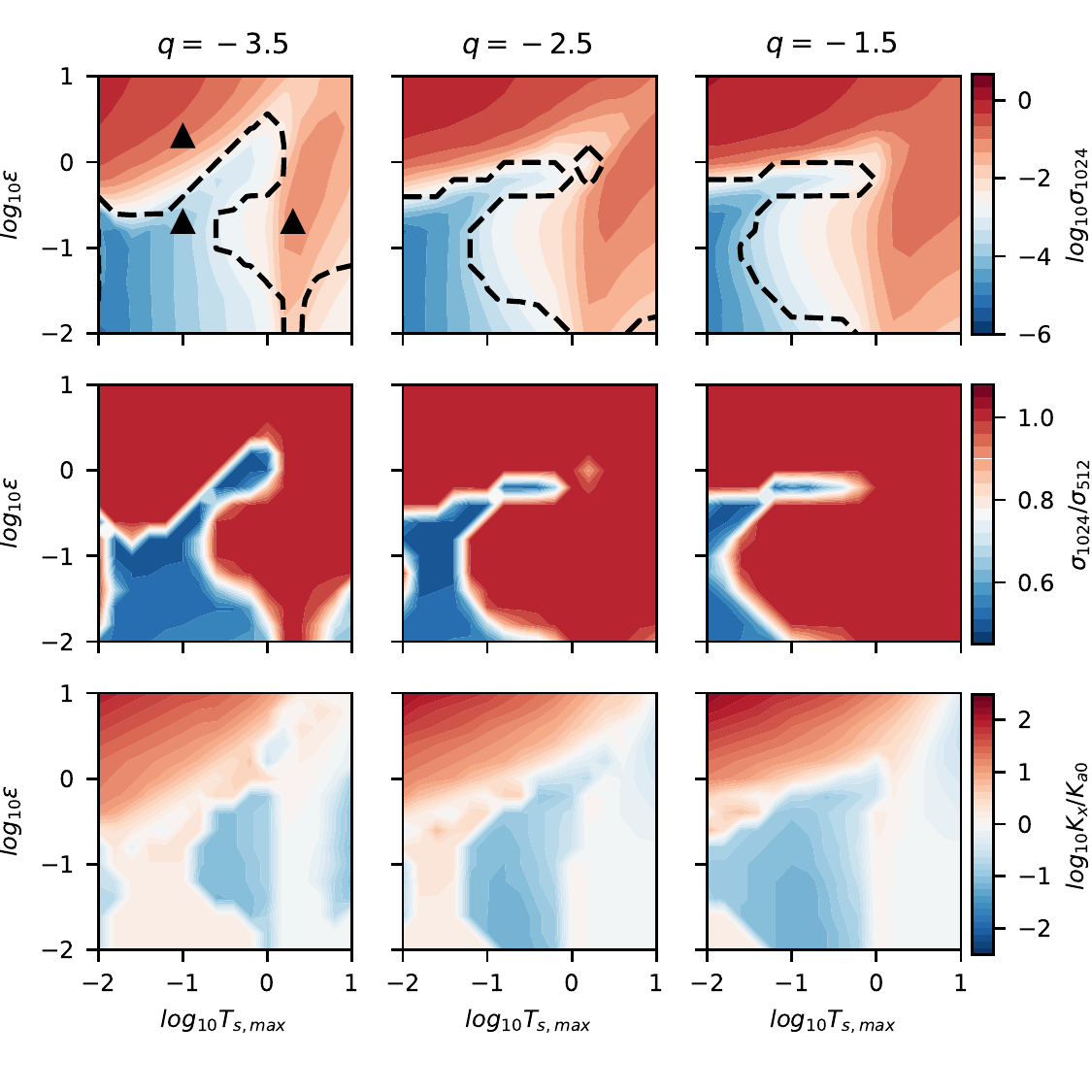}
\caption{%
    Similar to Fig.~\ref{fig:multi1} but with the minimum  {dust} size $\Tsmin = 10^{-3}$.
    The three triangles in the top-left panel denote the three cases which we have tested using direct numerical simulations (Sec.~\ref{SS:pc}).}
\label{fig:multimin1em3}
\end{figure*}

After studying the effects of $\Tsmax$ and $q$ on the instability growth rate, we explore the effects of $\Tsmin$ by increasing $\Tsmin$ from $10^{-4}$ to $10^{-3}$. The resulting growth rates are shown in Fig.~\ref{fig:multimin1em3}.
By comparing Fig.~\ref{fig:multimin1em3} with Fig.~\ref{fig:multi1}, we  {find} that the growth rate is not appreciably changed by $\Tsmin$ as long as $\Tsmax$ is not close to $\Tsmin$.
 {We investigate the special cases with $\Tsmax$ close to $\Tsmin$ in Sec.~\ref{S:para}.}
For the convenience of  {direct numerical simulations with which time steps are constrained} by the small dust species, we use $\Tsmin = 10^{-3}$ to validate the unstable modes in several representative cases in Fig.~\ref{fig:multimin1em3} (triangles in the figure),   {as presented in Sec.~\ref{SS:pc}}.

\subsection{Transition between the converged and non-converged regimes\label{SS:transition}}

As discovered by \cite{Krapp2019}, the fast and slow growth regimes discussed in Sec.~\ref{SS:cc} also seem to have distinct growth rate convergence with the number of discrete dust species $\Nsp$ representing the distribution.
 {Although we find that such connection between the fast (or slow) growth and converged (or non-converged) growth rate can break down for discs with a narrow dust size distribution (e.g.\ $\Tsmin \gtrsim 0.1\Tsmax$; see Sec.~\ref{S:para}), the distinctly different properties of convergence between the two regimes remain apparent for dust distributions with a wide size range.}
For $\Tsmin=10^{-3}$ and $10^{-4}$ cases, the growth rate in the fast-growth regime converges to a finite value, while in the slow-growth regime, the growth rate appears to approach zero, i.e., the system seems to be virtually stable to the streaming instability.
In this section, we attempt to identify the transition between these two regimes.

In the middle panels of Figs.~\ref{fig:multi1} and \ref{fig:multimin1em3}, we plot the ratio between $\sigma_{1024}$ and $\sigma_{512}$ for the same parameters as in the top panels.
This ratio decreases quickly from one (converged rates, which corresponds to the red colour) to $\sim$0.5 (blue colour) as either $\varepsilon$ or $\Tsmax$ decreases across some critical values.
The transition zone separating the two regimes appears to be excessively sharp, and thus we over-plot the contour of $\sigma_{1024} / \sigma_{512} = 0.99$ with dashed curves in the top panels to mark this transition.
If $\sigma_{2\Nsp} / \sigma_{\Nsp} \simeq 1$, the growth rate should approach to a finite value, while if $\sigma_{2\Nsp} / \sigma_{\Nsp} < 1$, the growth rate will continue to approach to zero as $\Nsp$ increases.
Therefore, with more and more dust species, we can expect that the growth rate remains the same in the upper and right sides of the dashed curve, while the growth rate below the dashed curve continues to decrease so that the blue region in the top panels ($\sigma$ panels) will become darker and darker, i.e., $\sigma_{\Nsp} \ll 1$.

The bottom panels in Figs.~\ref{fig:multi1} and~\ref{fig:multimin1em3} show the $K_x$ of the fastest growing mode. The $K_x$ value is normalized to $K_{a0}$. $K_{a0}$ represents
the $K_x$ of the fastest growth mode for a single dust species having $\Ts=\Tsmax$.
\cite{SH18} used the wave-drift resonance to show that the fastest growing short-wavelength mode for a single dust species with dust-to-gas mass ratio $\varepsilon \lesssim 1$ and stopping time $\Ts$ occurs at
\begin{equation}
    K_x=K_a\equiv\frac{(1+\varepsilon)^2+\Ts^2}{2(1+\varepsilon)\Ts}\,.\label{eq:Kx}
\end{equation}
When $\varepsilon\ll$1, $K_a$ is  insensitive to $\varepsilon$ so that we use $K_{a0}$, which is $K_a$ with $\varepsilon=0$, to scale $K_x$. We see that the converged and non-converged regimes have very different fastest growing modes,  {and we analyze them} in more detail in Sec.~\ref{S:wavenumber}.

\begin{figure*}
\includegraphics[trim=0mm 5mm 0mm 0mm, clip, width=6.in]{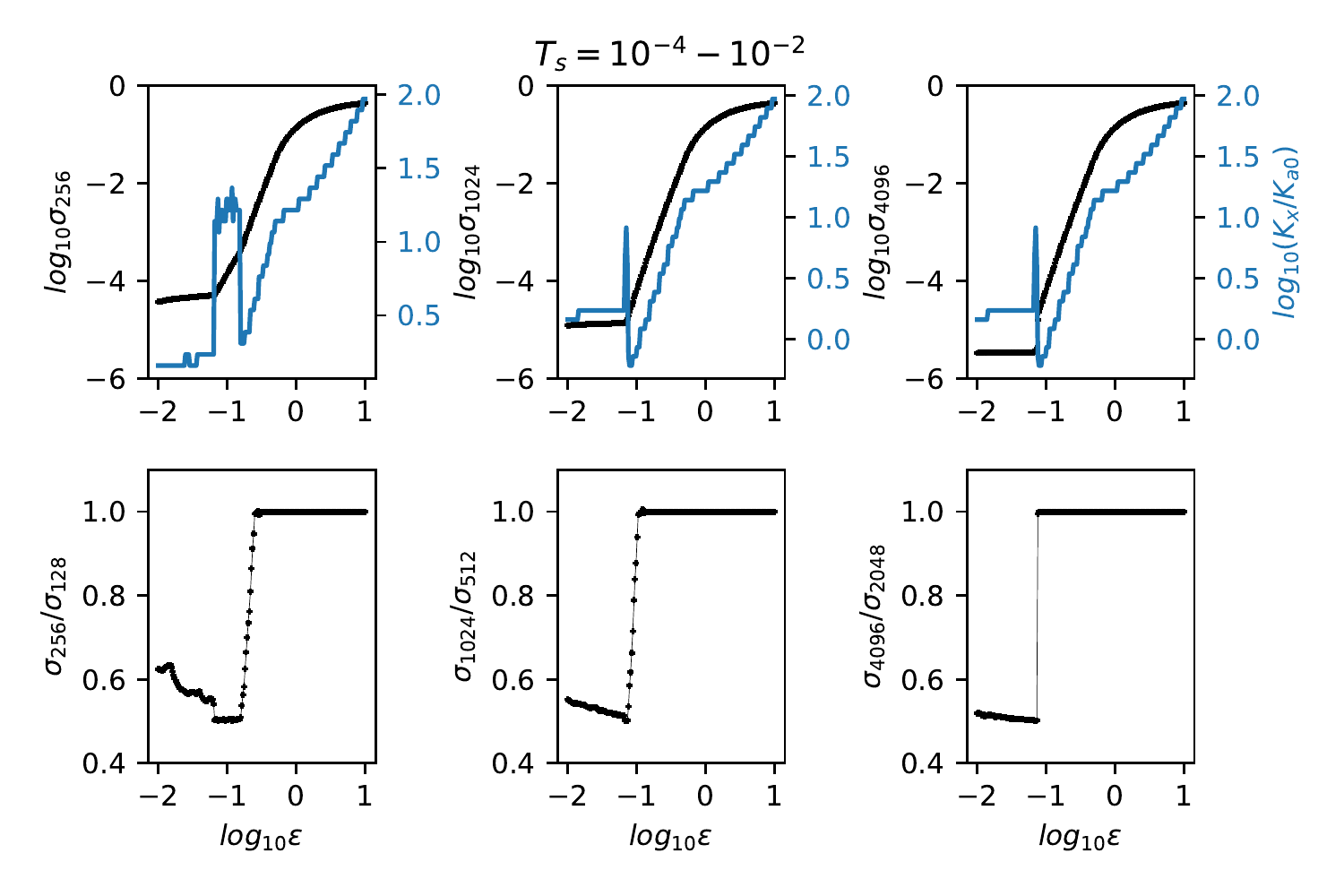}
\caption{%
    The growth rate $\sigma_{\Nsp}$ (in $\OmegaK$) and the dimensionless radial wave number $K_x$ of the fastest growing mode (top panels) and the growth rate ratio $\sigma_{2\Nsp} / \sigma_{\Nsp}$ (bottom panels) as a function of the total dust-to-gas mass ratio $\varepsilon$ for discs with a dust distribution of $\Ts \in [10^{-4}, 10^{-2}]$ and $q = -3.5$ (along the vertical dotted line in the top-left panel of Fig.~\ref{fig:multi1}).
    Number of dust species $\Nsp$ representing the distribution increases from the left to the right panels.
    \label{fig:ts0p01}}
\end{figure*}

\begin{figure*}
\includegraphics[trim=0mm 5mm 0mm 0mm, clip, width=6.in]{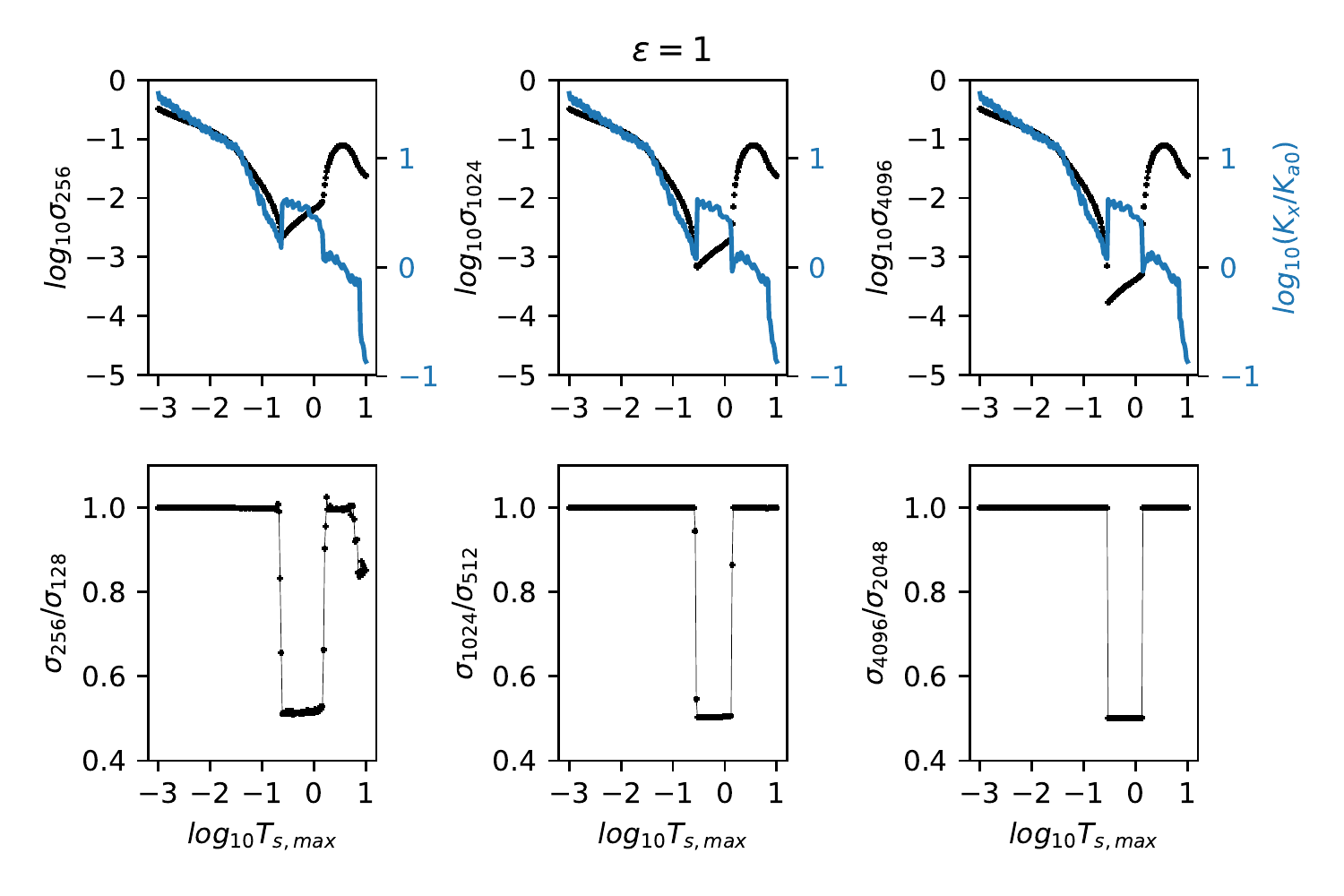}
\caption{%
    The growth rate $\sigma_{\Nsp}$ (in $\OmegaK$) and the dimensionless radial wave number $K_x$ of the fastest growing mode (upper panels) and the growth rate ratio $\sigma_{2\Nsp} / \sigma_{\Nsp}$ (bottom panels) as a function of the the maximum size $\Tsmax$ in a dust-size distribution with power-law index $q = -3.5$ (along the horizontal dotted line in the top-left panel of Fig.~\ref{fig:multi1}).
    The minimum size is $\Tsmin = 10^{-4}$ and the total dust-to-gas mass ratio is fixed at $\varepsilon = 1$.
    Number of dust species $\Nsp$ representing the distribution increases from the left to the right panels.
    \label{fig:epsilon1}}
\end{figure*}

To study how sharp this transition between the two regimes is, we conduct two additional sets of calculations.
For one set, we fix $\Tsmax = 0.01$ while varying $\varepsilon$ with 192 different values from 10$^{-2}$ to 10.
For the other set, we fix $\varepsilon = 1$ while varying $\Tsmax$ with 192 different values from $10^{-3}$ to 10.
For both sets, we consider only the dust-size distribution with power-law index $q = -3.5$.
The trace of these two fine scans are schematically drawn with the two dotted lines in the top-left panel of Fig.~\ref{fig:multi1}, and the resulting maximum growth rates are shown in Figs.~\ref{fig:ts0p01} and~\ref{fig:epsilon1}, respectively. The number of dust species $\Nsp$ increases from 256 in the left panels to 4096 in the right panels.%
\footnote{ {We caution that the growth rates for 2048 and 4096 species are calculated at the same wave number ($K_x$, $K_z$) of the fastest growing mode for 1024 species (Sec.~\ref{SS:la}).}}

As shown by the bottom panels of Fig.~\ref{fig:ts0p01}, the growth rate converges with number of dust species to finite values ($\sigma_{2\Nsp}/\sigma_{\Nsp} = 1$) for $\Ts \in [10^{-4}, 10^{-2}]$ and $\varepsilon \gtrsim 0.1$.
The transition from $\sigma_{2\Nsp}/\sigma_{\Nsp} \sim 0.5$ to $\sigma_{2\Nsp}/\sigma_{\Nsp} = 1$ becomes sharper and sharper with increasing $\Nsp$, which can also be seen in the upper panels.
While the growth rate for $\varepsilon \gtrsim 0.1$ remain the same from the left to the right top panels (increasing $\Nsp$), the growth rate for $\varepsilon \lesssim 0.1$ becomes smaller and smaller.
This leads to a discontinuity of the growth rate which can be identified as the boundary between the two regimes.
With larger number of dust species, the fast-growth regime remains to have high growth rate while the slow-growth regime has decreasingly low growth rate.

We now consider the  {systems} with a fixed $\varepsilon = 1$ and different $\Tsmax$, along the horizontal dotted line in the top-left panel of Fig.~\ref{fig:multi1}.
The growth rate as a function of $\Tsmax$ is shown in the top panels of Fig.~\ref{fig:epsilon1} with various $\Nsp$.
For $\Tsmax \lesssim 0.02$ and $\Tsmax \gtrsim 0.2$, the profile of the growth rate remains the same with increasing $\Nsp$, as also evident in the bottom panels where $\sigma_{2\Nsp}/\sigma_{\Nsp} = 1$.
For $0.02 \lesssim \Tsmax \lesssim 0.2$, on the contrary, the rate becomes lower and lower with increasing $\Nsp$, leading to a growth rate orders of magnitude lower than in the fast-growth regime when $\Nsp = 4096$.
Similar to the case of fixed dust-size distribution but varying $\varepsilon$, the transitions at $\Tsmax \simeq 0.02$ and $\Tsmax \simeq 0.2$ become sharper and sharper with increasing $\Nsp$.
Therefore, the two regimes appear to be distinctly separated by a well defined boundary, as delineated by the dashed lines in Figs.~\ref{fig:multi1} and~\ref{fig:multimin1em3}.

\subsection{Numerical validation\label{SS:pc}}

\begin{table}
	\centering
	\caption{Growth rate $\sigma$ (in $\OmegaK$) of fast-growing eigenmode in multi-species streaming instability to be compared with direct numerical simulations}
	\label{tab:sigmatable}
	\begin{tabular}{rccc} 
		\hline
		       & $\Ts \in [10^{-3}, 0.1]$ & $\Ts \in [10^{-3}, 0.1]$ & $\Ts \in [10^{-3}, 2]$\\
		       & $\varepsilon = 2$        & $\varepsilon = 0.2$      & $\varepsilon = 0.2$\\
		$\Nsp$ & $K_x = K_z = 60$         & $K_x = K_z = 10$         & $K_x = K_z = 1$ \\
		\hline
		  1    & 0.400922\phn             & 0.00748144               & 0.0885570\\
		  2    & 0.0616073                & 0.00787742               & 0.0857477\\
		  4    & 0.169425\phn             & 0.00615058               & 0.0794737\\
		  8    & 0.127082\phn             & 0.00334111               & 0.0519613\\
		 16    & 0.0976177                & 0.00283122               & 0.0573683\\
		 32    & 0.0980250                & 0.00169562               & 0.0450016\\
		 64    &                          &                          & 0.0386043\\
		128    &                          &                          & 0.0397249\\
		\hline
	\end{tabular}
\end{table}
As described in Sec.~\ref{SS:nv}, we use the \pc{} to mutually validate the eigenmodes derived from our linear analysis.
We select three different dust distributions:
(1)~$\Ts \in [10^{-3}, 0.1]$ and $\varepsilon = 2$,
(2)~$\Ts \in [10^{-3}, 0.1]$ and $\varepsilon = 0.2$, and
(3)~$\Ts \in [10^{-3}, 2]$ and $\varepsilon = 0.2$,
all of which have a power-law index of $q = -3.5$.
For each distribution, we choose an eigenmode with a wave number $K_x = K_z$ that is close to the fastest growing mode.
The three cases are marked by the triangles in Fig.~\ref{fig:multimin1em3}, and the growth rate of the mode as a function of number of dust species $\Nsp$ is listed in Table~\ref{tab:sigmatable}.

\begin{figure*}
\includegraphics[width=0.8\textwidth]{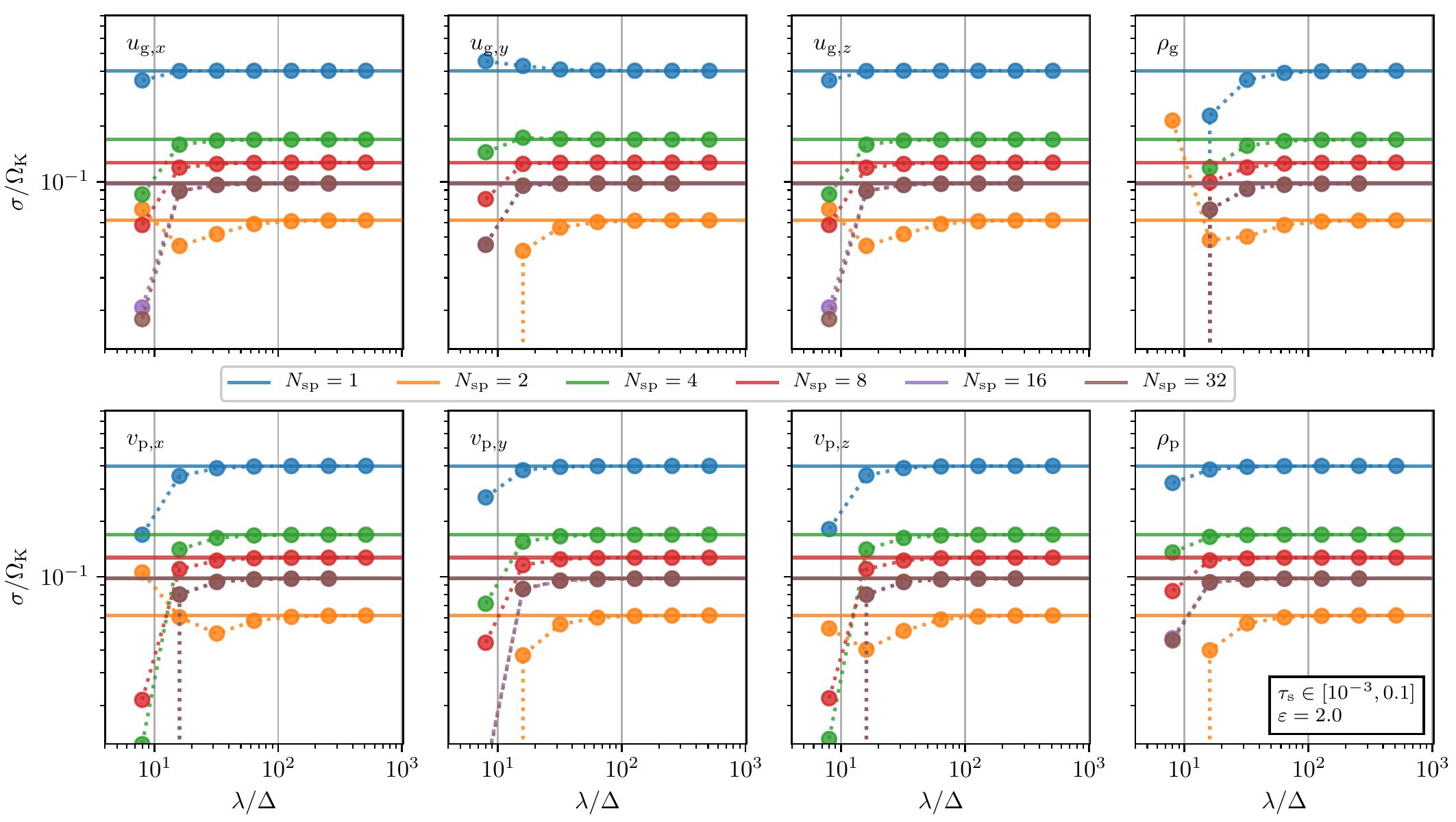}
\caption{Numerical validation of linear modes of the streaming instability for different numbers of dust species using the \pc.
    The dots are the growth rate $\sigma$ of a mode in units of $\OmegaK$, measured from the simulation data at various resolution in terms of number of grid points per wavelength $\lambda / \Delta$.
    Different colours represent different numbers of dust species $\Nsp$ in the system.
    The solid lines are the theoretical growth rates.
    The top panels show the components of the velocity field and the density of the gas, while the bottom panels show the components of the mass-weighted velocity field and the total density of the dust particles.
    In this case, the dimensionless stopping time is in the range of $\Ts \in [10^{-3}, 0.1]$, and the total solid-to-gas mass ratio is $\varepsilon = 2$.
    The selected mode has the wave number $K_x = K_z = 60$, which is close to the fastest growing mode.
    We note that the results are close to identical between the $\Nsp = 16$ and the $\Nsp = 32$ cases.
    \label{fig:pc0120}}
\end{figure*}

First, we consider the distribution $\Ts \in [10^{-3}, 0.1]$ and $\varepsilon = 2$.
As shown in Fig.~\ref{fig:multimin1em3}, this case belongs to the converged fast-growth regime.
The growth rate approaches to a constant with increasing number of dust species and convergence occurs at $\Nsp \sim 16$, as indicated by Table~\ref{tab:sigmatable}.
Figure~\ref{fig:pc0120} shows the growth rate measured by the \pc{} with various $\Nsp$ and resolutions $\lambda / \Delta$, where $\lambda = 2\pi / k_i$ is the wavelength per dimension with $i = x, z$ and $\Delta$ is the grid spacing.
The simulation data demonstrate convergence with resolution to the theoretical rate at each given $\Nsp$.
For $\Nsp > 2$, convergence of gas velocity and dust density is seen at $\lambda / \Delta \sim 32$ while convergence of dust velocity and gas density is seen at $\lambda / \Delta \sim 64$.
For the case of $\Nsp = 1$, the convergence with resolution is similar while it is slightly worse for the case of $\Nsp = 2$.

\begin{figure*}f
\includegraphics[width=0.8\textwidth]{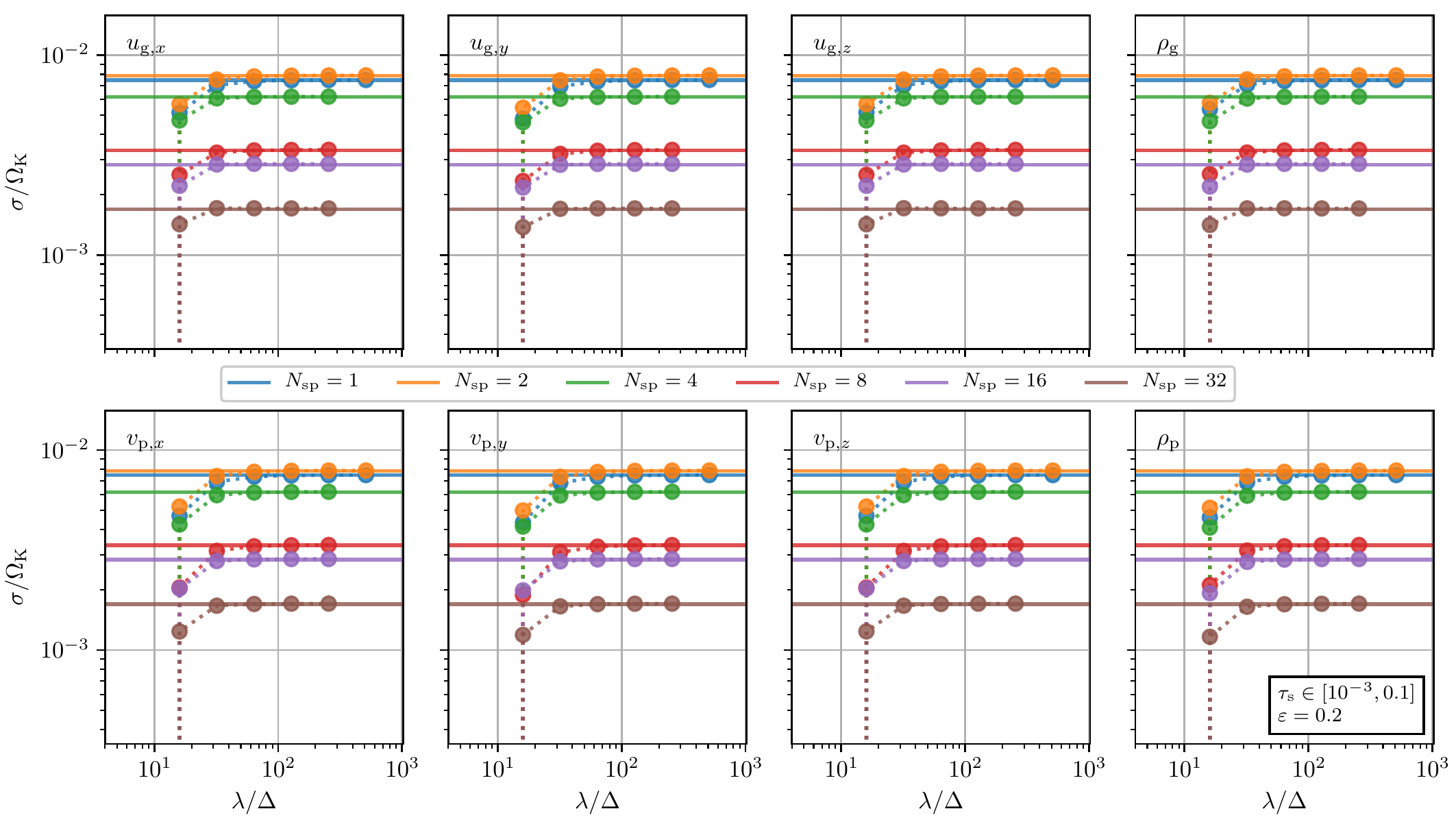}
\caption{Same as Fig.~\ref{fig:pc0120}, except that $\varepsilon = 0.2$ and $K_x = K_z = 10$.\label{fig:pc0102}}
\end{figure*}

We next consider the dust distribution $\Ts \in [10^{-3}, 0.1]$ and $\varepsilon = 0.2$.
It only differs from the previous case by the much decreased solid-to-gas mass ratio $\varepsilon$, leading to longer critical wavelengths and into the non-converged slow-growth regime (Fig.~\ref{fig:multimin1em3}).
As shown by Table~\ref{tab:sigmatable}, the growth rate continues to decrease with increasing number of dust species $\Nsp$.
Again we use the \pc{} to measure the growth rate up to $\Nsp = 32$ with varying resolutions, and the results are shown in Fig.~\ref{fig:pc0102}.
The code can still capture the instability accurately at such a low growth rate of $\sigma \simeq 0.0017\OmegaK$ and convergence with resolution can be seen at $\lambda / \Delta \sim 64$.
For comparison, the LinB mode for the single-species streaming instability ($\Ts = 0.1$ and $\varepsilon = 0.2$) has a growth rate of $\sigma \simeq 0.015\OmegaK$, the lowest ever used for code validation in the previous literature \citep{YJ07,BT09,BS10,fM10,YJ16,Benitez2019,MFV19, Krapp2019}.
The case we validate here has a growth rate one order of magnitude lower and hence is more challenging.

\begin{figure*}
\includegraphics[width=0.8\textwidth]{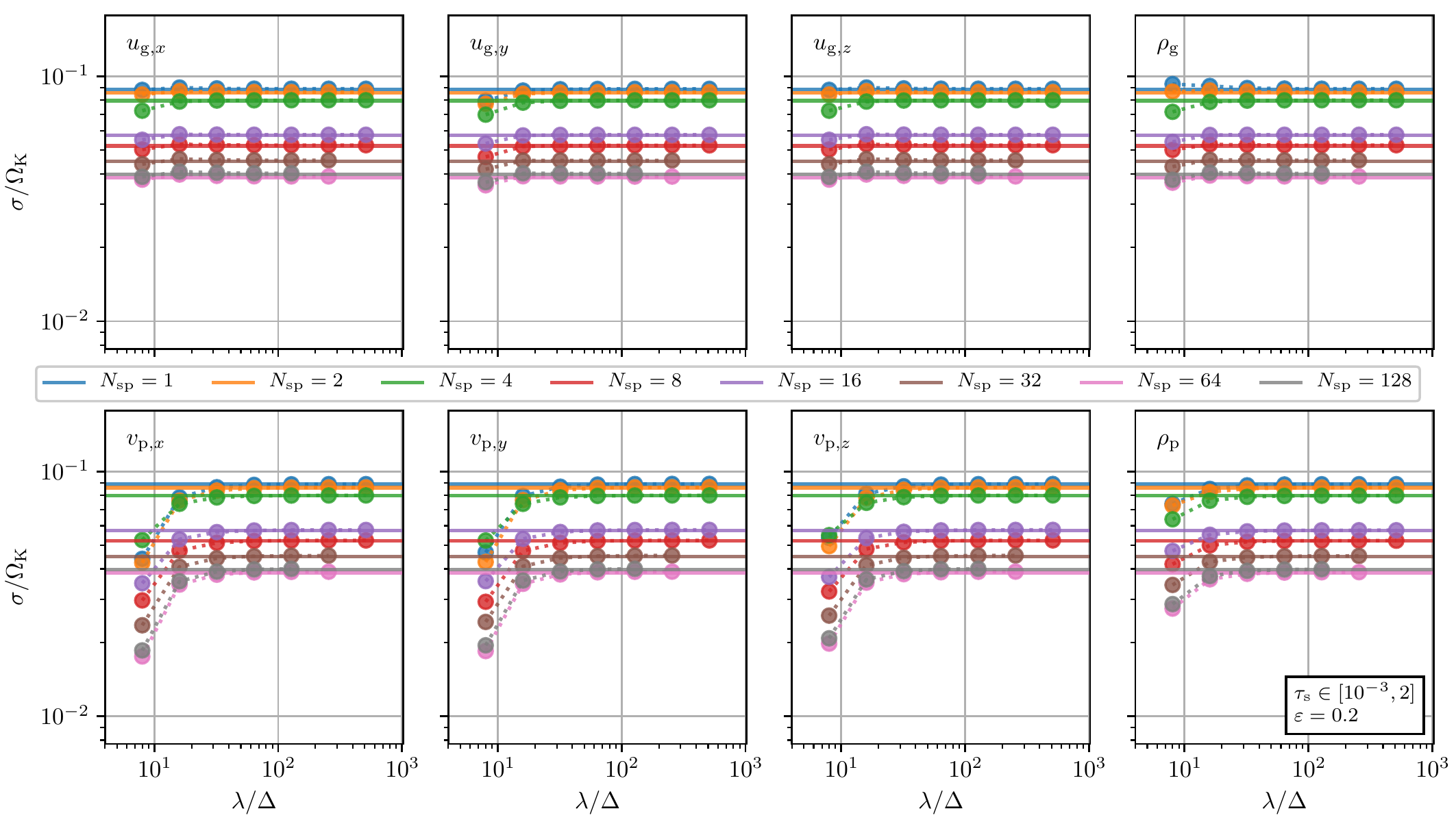}
\caption{Same as Fig.~\ref{fig:pc0120}, except that $\Ts \in [10^{-3}, 2]$, $\varepsilon = 0.2$ and $K_x = K_z = 1$.\label{fig:pc2001}}
\end{figure*}

Finally, we change the maximum size of the dust distribution from the previous case to $\Tsmax = 2$, the latter of which is in the converged fast-growth regime (Fig.~\ref{fig:multimin1em3}).
This case is valuable in the sense that the largest dust particles probe into the regime of $\Ts \sim 1$, above which the fluid approximation used by the linear analysis may be broken because the trajectories of the particles may begin to cross \citep{GBL04}.
The growth rate measured with the \pc{} as a function of the number of dust species $\Nsp$ and the resolution $\lambda / \Delta$ is shown in Fig.~\ref{fig:pc2001}.
Once again, convergence of the growth rate to the theoretical rate with resolution at any given $\Nsp$ is achieved.
The growth in the velocity and density field of the gas converges at $\lambda / \Delta \sim 16$--32, while that of the dust particles converges at $\lambda / \Delta \sim 32$--64.
This experiment indicates that the fluid approximation for the dust particles is still valid in the linear growth of the instability for the largest particles up to $\Tsmax \sim 2$.
However, it remains to be seen if the approximation holds valid in the nonlinear saturation of the instability.

\section{Discussion}
\subsection{On the critical wave number} \label{S:wavenumber}

To explore further the nature of the instability in the fast/slow  {or} converged/non-converged regimes, we turn our attention to the most unstable mode itself for any given system.
Based on the linear analysis of the single-species streaming instability, \cite{YoudinGoodman2005} found that the most unstable mode occurs along $K_z \sim \Ts K_x^2$ on the long-wavelength side and $K_x \sim \mathrm{constant}$ on the short-wavelength side, which are bridged roughly at $K_x \sim K_z \sim 1 / \Ts$.
Usually, the maximum growth rate can be found near $K_x \sim K_z \sim 1 / \Ts$ or on the vertical $K_x \sim 1 / \Ts$.
This property is closely related with the resonant drag instability, in which the relative drift velocity between the gas and the dust resonates with the projected wave speed of the gas \citep{SH18}.
With this new insight into the streaming instability, \cite{SH18} gave a physically motivated $K_x$--$K_z$ condition for the long-wavelength branch (their equation~(33)) and $K_x = K_a$ (equation~\eqref{eq:Kx}) for the short-wavelength branch, assuming $\varepsilon \lesssim 1$ \citep[see also][]{UEC20}.
In the following, we use it as a reference to analyze what we find in the multi-species streaming instability.

The  {bottom} panels in Figs.~\ref{fig:multi1} and \ref{fig:multimin1em3} show $K_x$ of the mode that has the highest growth rate.
The dimensionless radial wave number $K_x$ is normalized by $K_{a0}$.
Therefore, if the fastest growing mode is near the vertical branch $K_x = K_{a0}$, the colour should be white in these panels, and it can be implied that the largest dust particles drive the instability in the white region.
As shown by the panels, the non-convergent cases (which correspond to the blue region in the panels in the middle row) roughly have $K_x / K_{a0} \sim 1$--3.
The convergent cases can be divided by $\varepsilon \sim 1$.
For $\varepsilon\gtrsim 1$, $K_x\gtrsim 10 K_{a0}$.
For $\varepsilon\lesssim 1$, $K_x$ depends on the maximum stopping time $\Tsmax$:
When $\Tsmax$ is small, the fast growing mode has $K_x\ll K_{a0}$ (the vertical wave number $K_z$ is similarly small such that the modes may not fit within the disc thickness, as discussed below), and when $\Tsmax\gtrsim1$, $K_x\sim K_{a0}$.

As in Sec.~\ref{SS:transition}, we perform a vertical cut in the bottom-left panel of Fig.~\ref{fig:multi1} as in the top-left panel, and plot $K_x$ of the fastest growing mode (blue curves in Fig.~\ref{fig:ts0p01}) as a function of $\varepsilon$ with a fixed dust distribution $\Ts \in [10^{-4}, 10^{-2}]$ and $q = -3.5$.
Similar to the convergence of the growth rate, we notice that $K_x$ also has a dramatic change at the $\varepsilon\sim0.1$ convergence boundary, indicating that non-convergent and convergent cases may have different types of the fastest growing modes.
When $\varepsilon\lesssim0.1$, $K_x$ is close to 1--2$K_{a0}$.
At $\varepsilon\sim0.1$, $K_x$ jumps to 20$K_{a0}$ for 256 dust species; for 1024 and 4096 dust species, there is only a narrow $\varepsilon$ space for these high $K_x$ values.
For convergent cases with $\varepsilon \gtrsim 0.1$, $K_{x}$ drops to $\sim K_{a0}$ and then increases with larger $\varepsilon$.

\begin{figure*}
\includegraphics[trim=0mm 2mm 0mm 0mm, clip, width=6.6in]{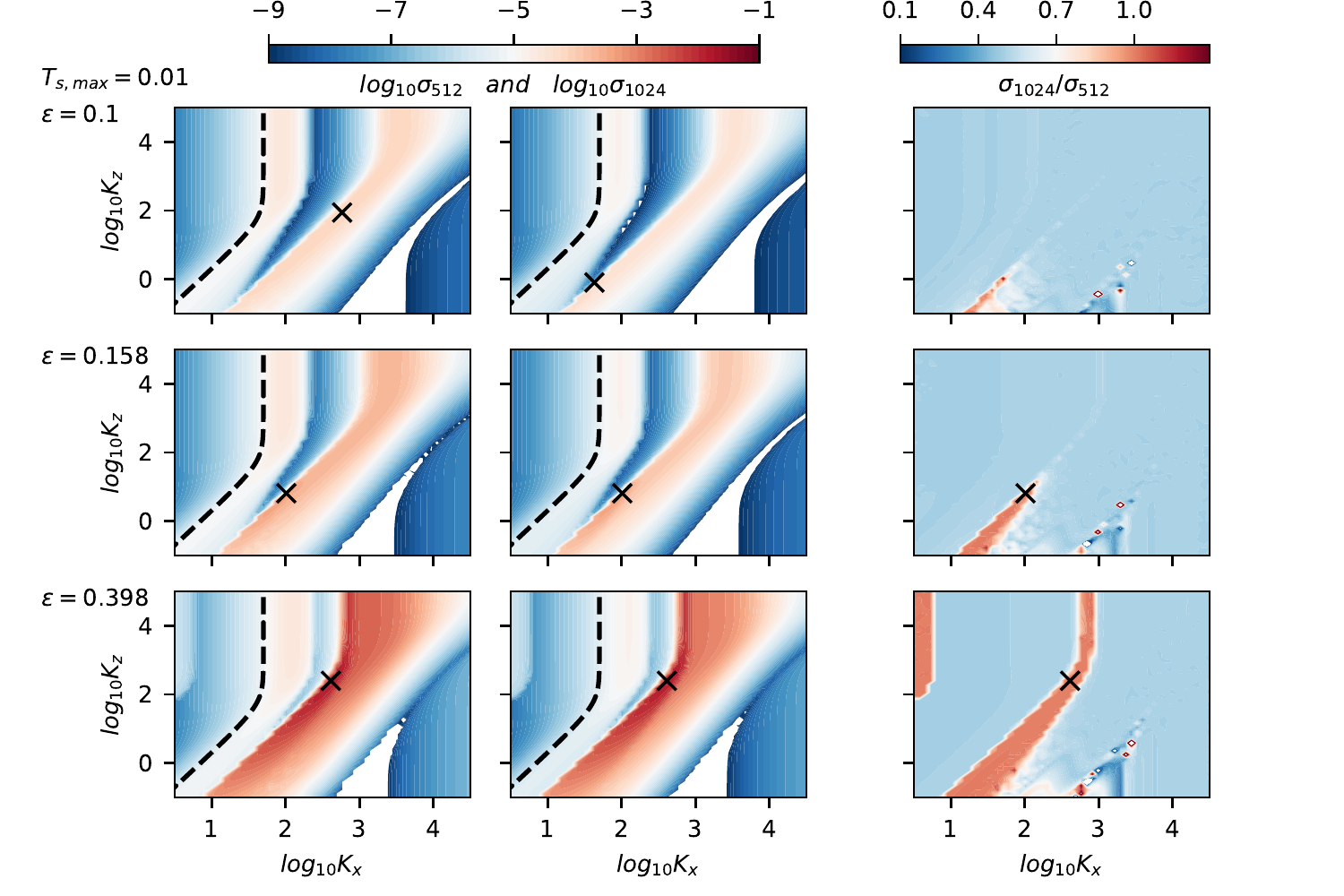}
\caption{
     Growth rate of the most unstable mode $\sigma$ (in $\OmegaK$) as a function of the dimensionless wave number ($K_{x}$, $K_{z}$).
    The discs have the same dust distribution $\Ts \in [10^{-4}, 10^{-2}]$ and $q = -3.5$, but the total solid-to-gas mass ratio $\varepsilon$ increases from the top to the bottom panels.
    The three cases are represented by the triangles in the top-left panel of Fig.~\ref{fig:multi1}.
    The left panels show the growth rate for discs with 512 dust species, while the middle panels show the growth rate for discs with 1024 dust species.
    The right panels show the ratio between the left two panels.
    If the ratio at one point is one, the growth rate of that mode converges  {to finite values}.
    The dashed curves in the left two columns are calculated using equation~(33) in Squire \& Hopkins (2018) near which the system with a single dust species having $\varepsilon$ and $\Ts = \Tsmax$ is the most unstable.   The cross in each panel labels where the fastest growing mode is.  \label{fig:test}}
\end{figure*}

To study these modes in detail, we plot in Fig.~\ref{fig:test} the growth rate with respect to $K_x$ and $K_z$ for three cases with
the same $\Tsmax$ but different $\varepsilon$ (located at the triangles in the top-left panel of Fig.~\ref{fig:multi1}).
It can be seen that the fast growing modes (red coloured) may lie in two separate regions: the region that is close to the dust-gas drag resonance with single dust species $\Ts = \Tsmax$ \citep[dashed curves;][]{YoudinGoodman2005,SH18}, and another region whose $K_x$ is about one order of magnitude larger (which was also pointed out by \citealt{Krapp2019}).
When $\varepsilon$ is as small as $\sim$0.01 (not shown), the fastest growing mode lies in the region that is close to the dashed curves.
With larger $\varepsilon\sim0.1$ (the uppermost panels of Fig.~\ref{fig:test}), the most unstable mode lies in the larger $K_x$ region.
As shown in the upper right panel, both red regions do not show convergent rates with increasing $\Nsp$.
Thus, the instability growth rate does not converge to finite values when $\varepsilon\lesssim0.1$, consistent with Fig.~\ref{fig:ts0p01}.
When $\varepsilon$ continues to rise (the middle row in Fig.~\ref{fig:test}), the narrow region with $K_z\sim \Tsmax K_x^2/10$
and small $K_x$ starts to show convergence and hence this region maintains its high growth rate with increasing $\Nsp$ (left and middle panels).
This converged region extends to higher $K_x$ and $K_z$ when $\varepsilon$ is even larger (the bottom panels).
These changes of the convergent regions with $\varepsilon$ explain the trend of $K_x$ shown in Fig.~\ref{fig:ts0p01}.

\begin{figure*}
\includegraphics[trim=0mm 2mm 0mm 0mm, clip, width=6.6in]{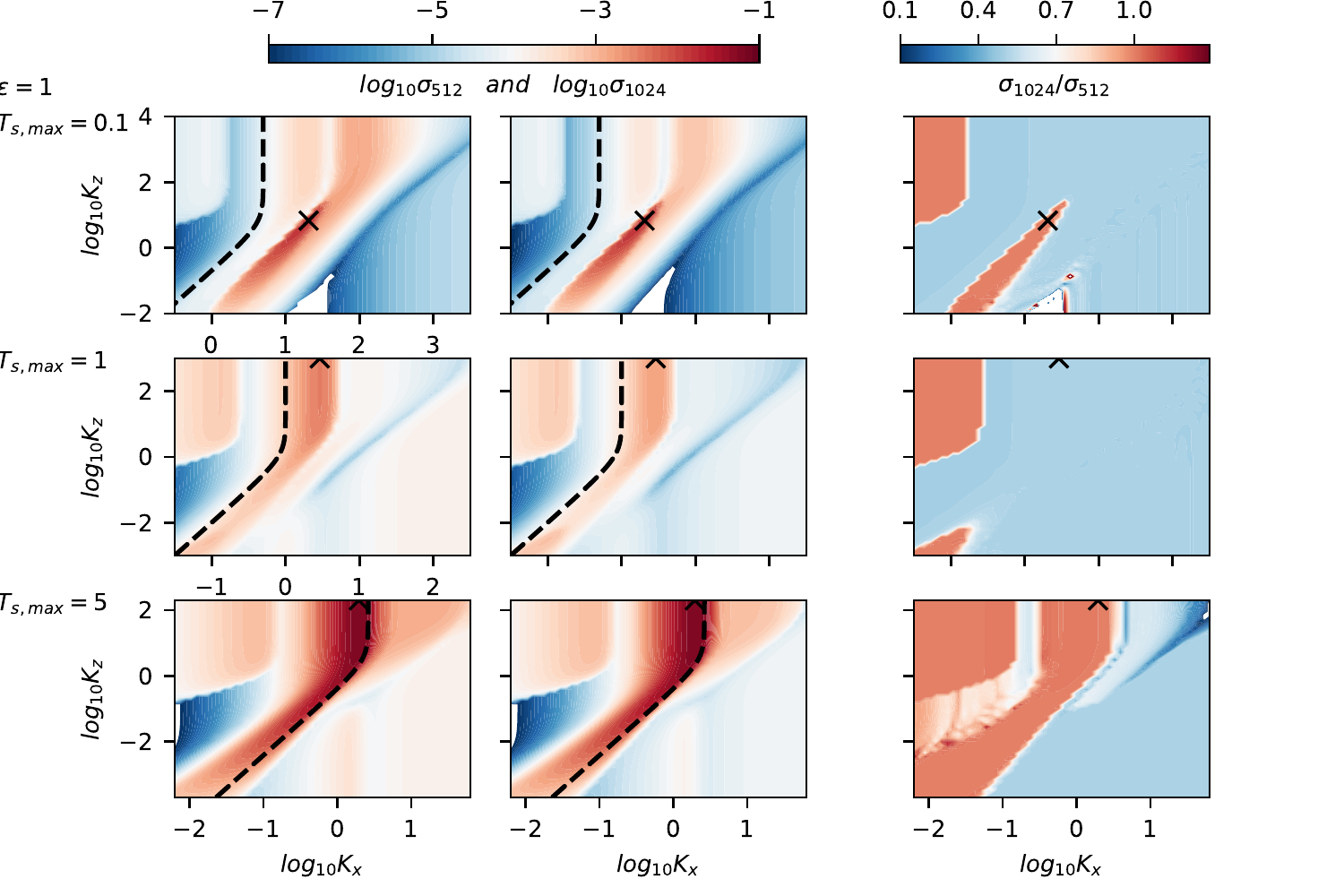}
\caption{%
    Similar to Fig.~\ref{fig:test} but for discs with the same $\varepsilon = 1$ and $\Tsmin = 10^{-4}$ but varying $\Tsmax$.
    The value of $\Tsmax$ increases from the top to the bottom panels, and these cases are located at the three crosses in the top-left panel of Fig.~\ref{fig:multi1}.
    \label{fig:test2}}
\end{figure*}
We next consider the horizontal cut in the upper-left panel of Fig.~\ref{fig:multi1}, i.e., dust distributions with fixed $\varepsilon = 1$, $q = -3.5$, and $\Tsmin = 10^{-4}$ but with varying $\Tsmax$.
The blue curves in the top panels of Fig.~\ref{fig:epsilon1} show the radial wave number $K_x$ of the fastest growing mode, and Fig.~\ref{fig:test2} shows the map of the growth rate with respect to $K_x$ and $K_z$ for three selected cases (located at the crosses in Fig.~\ref{fig:multi1}).
For cases shown in both the top and the bottom panels of Fig.~\ref{fig:test2}, the growth rate converges, while the growth rate does not converge for the case in the middle panels.
The top panels show that when $\Tsmax \ll 1$, the fastest growing mode is at the large $K_x$ branch.
When $\Tsmax$ increases to $\sim$1, the convergent region shrinks significantly (right panels), and the fastest growing mode occurs in the single-species resonant region $K_x \sim K_{a0}$ which does not show convergence with $\Nsp$.
When $\Tsmax$ increases to even larger values (e.g.\ $\Tsmax = 5$; the bottom panels), the fastest growing mode remains in this $K_x \sim K_{a0}$ region, but the region becomes converged with $\Nsp$.

Figures~\ref{fig:test} and~\ref{fig:test2} show that unstable modes have converged growth rates with increasing $\Nsp$ in only a small region of the $K_x$--$K_z$ space.
If the fastest growing mode locates in this region, the growth rate of the instability converges to finite values with $\Nsp$.
The $K_x\sim K_{a0}$ region where the single-species streaming instability has the fastest growth rates sometimes does not show convergence with $\Nsp$.
Instead, a different region with larger $K_x$ can converge under some $\Tsmax$ and $\varepsilon$ conditions.
For future analytical studies, it will be crucial to understand why the growth rate of unstable modes at any given $K_x$ and $K_z$ do or do not converge with $\Nsp$.

\begin{figure*}
\includegraphics[trim=0mm 3mm 0mm 0mm, clip, width=6.in]{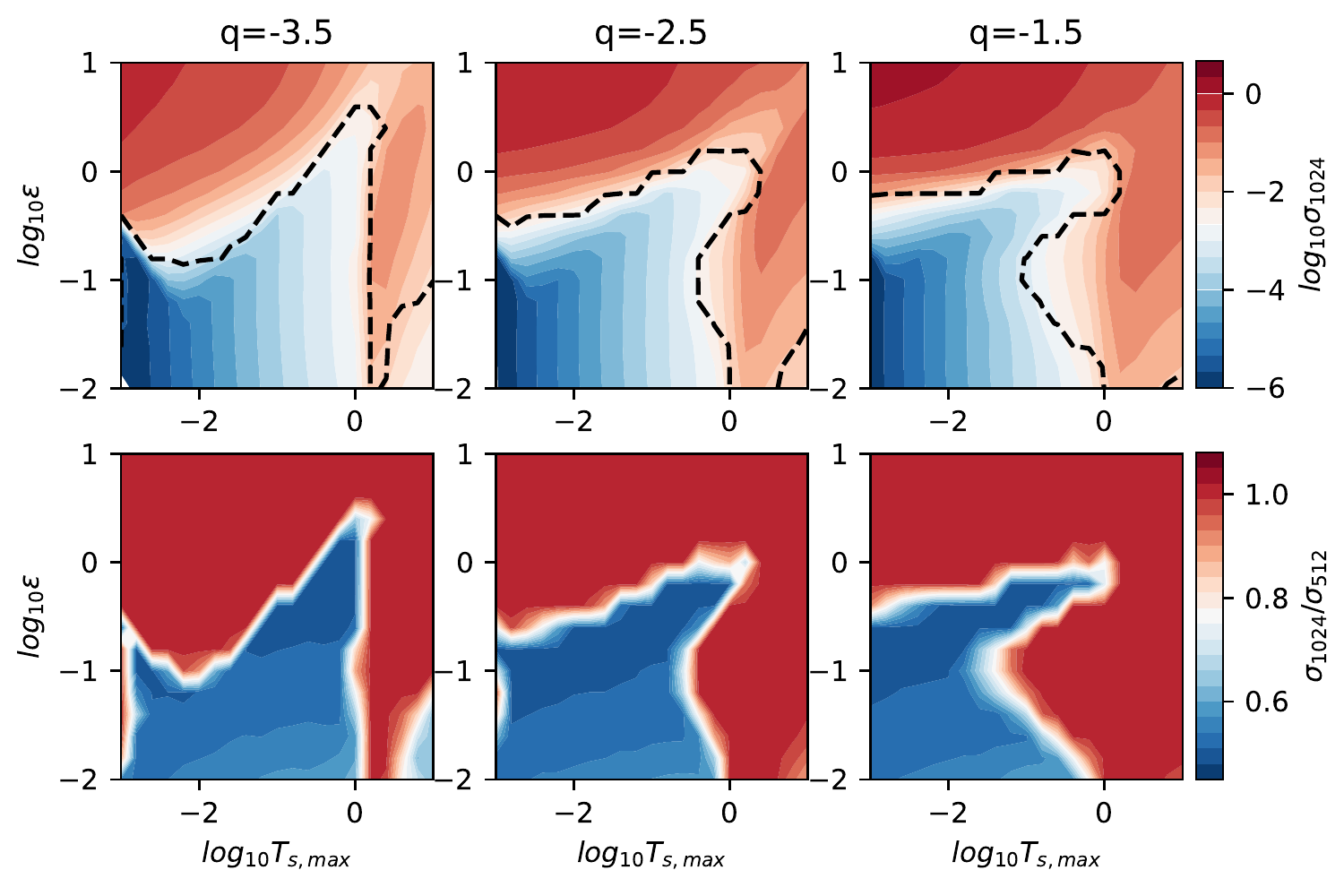}
\caption{Similar to Fig.~\ref{fig:multi1} but only the unstable modes with $k_z\geqslant 1/H_0$ are considered.}
\label{fig:multi2}
\end{figure*}
Finally, we note that when the vertical wave number of the most unstable mode $k_z\lesssim 1/H_0$, the wavelength of the mode is larger than the disc thickness and thus the instability may not operate in the disc.
Since $K_z = k_z\eta R_0$ and $\eta v_K/c_s=0.05$, $K_z$ of the unstable mode needs to be larger than 0.05 to fit into the disc.
Thus, in Fig.~\ref{fig:multi2}, we only include modes with $K_z \geqslant 0.05$ for the computation of the maximum growth rate.
Compared with Fig.~\ref{fig:multi1}, it can be seen that the converged region shrinks around $\Tsmax\sim 1$ and $\varepsilon \lesssim 1$, and the boundary between the convergent and the non-convergent regions is more aligned with $\Tsmax\sim 1$.
This difference can be understood by inspecting the $K_x$ values of the most unstable mode in the lower panels of Fig.~\ref{fig:multi1}, where $K_x$ can become much less than $K_{a0}$ around $\Tsmax\sim1$.
Since $K_z\sim K_x$ for these fastest growing modes, this implies that $K_z$ can be so small around $\Tsmax\sim1$ that the modes can be too large to fit in the disc.

\subsection{Interaction between Dust Species} \label{S:interaction}
\cite{Krapp2019} showed that when the total solid-to-gas density ratio $\varepsilon \ll 1$ for a given dust-size distribution, the instability is as if being driven by the largest dust species alone.
Therefore, as more and more discrete species $\Nsp$ are involved, the mass fraction of the leading species becomes smaller and smaller, resulting in increasingly small growth rate with increasing $\Nsp$ and hence its non-convergence.
Counter-intuitively, as $\varepsilon$ becomes significant, the growth of the instability with multiple species is even slower than the leading species in isolation would have driven.
In this section, we use a different approach to explore this property in a different angle and attempt to gain some more insight into the role of the leading dust species and the interaction between the species.

Our experiment is designed as follows.
As in Sec.~\ref{SS:la}, we divide a given dust-size distribution $\Ts \in [10^{-4}, \Tsmax]$ with total solid-to-gas density ratio $\varepsilon$ into $\Nsp$ regular logarithmic bins, with either $\Nsp = 512$ or $\Nsp = 1024$.
Each bin then has a solid-to-gas density ratio $\varepsilon_j$ (equation~\eqref{E:epsj}) and is represented by identical dust particles with dimensionless stopping time $\Tsi{j}$, where $\Tsi{1} < \Tsi{2} < \cdots < \Tsi{\Nsp}$.
We begin with the bin with the largest dust species $\Tsi{\Nsp}$ and find the fastest growing mode assuming that it is the only species present in the distribution.
Then, we activate the bin with the second largest dust species $\Tsi{\Nsp - 1}$ in the distribution and find the fastest growing mode again for this case.
In this manner, we systematically add more and more smaller dust species until we recover the full distribution.

\begin{figure*}
\includegraphics[trim=0mm 0mm 0mm 0mm, clip, width=6.6in]{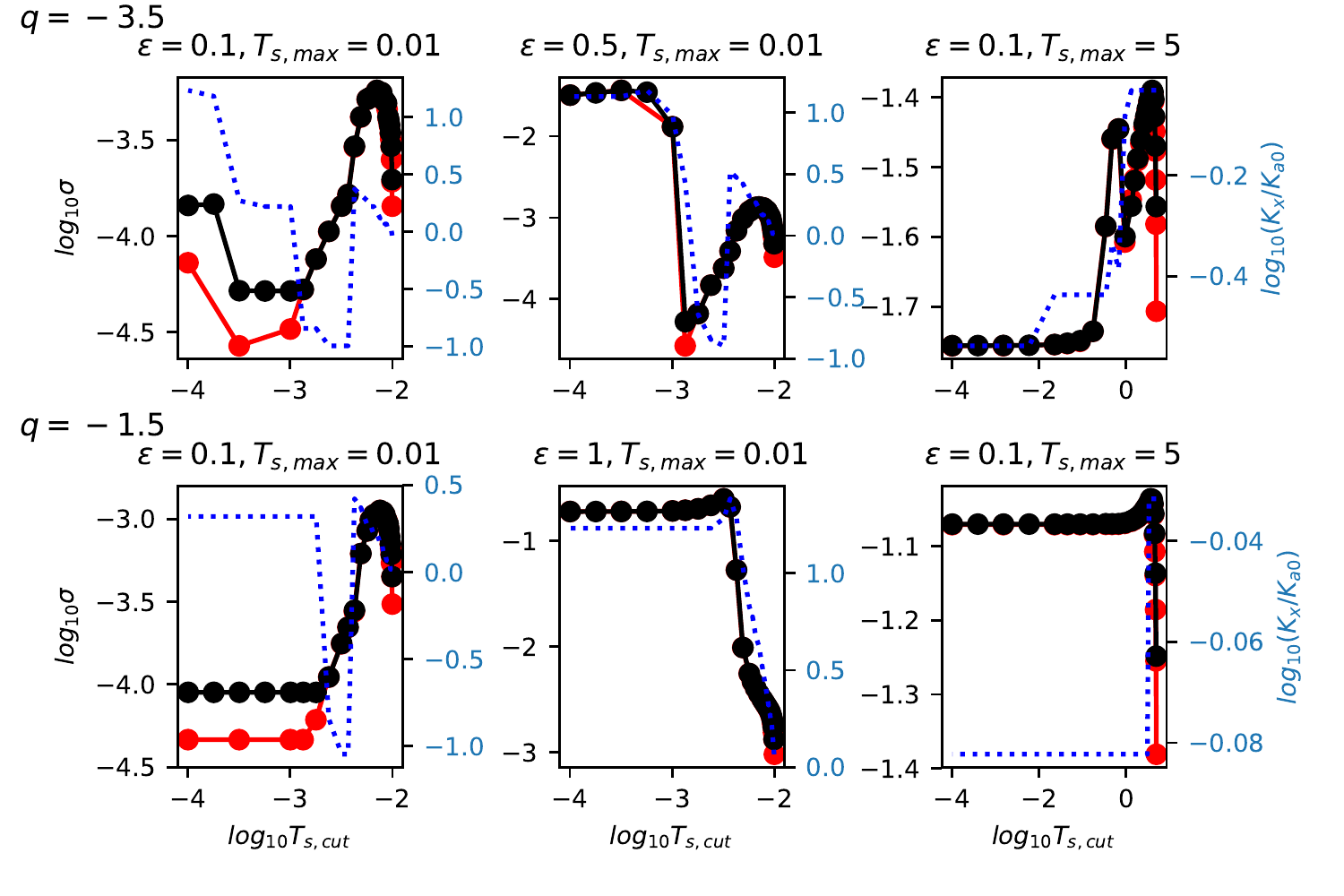}
\caption{%
    Maximum growth rate $\sigma$ (dots; left axis) and dimensionless radial wave number $K_x$ (dotted lines; right axis) as a function of the lower cutoff from a given dust-size distribution.
    Each panel represents a distribution with given dimensionless stopping time $\Ts \in [10^{-4}, \Tsmax]$, power-law index $q$, and \emph{total} solid-to-gas ratio $\varepsilon$, and the horizontal axis is the minimum stopping time $T_\mathrm{s,cut}$ cut from such a distribution.
    The upper panels and the lower panels have $q = -3.5$ and $q = -1.5$, respectively.
    The black dots mean the \emph{full} distribution contains 512 dust species, while the red dots 1024 species.
    The blue curves are the $K_x$ of the fastest mode with 512 dust species.
    \label{fig:increasespe1}}
\end{figure*}
The results for various $\Tsmax$, $\varepsilon$, and $q$ are shown in Fig.~\ref{fig:increasespe1}.
Comparing the black dots ($\Nsp = 512$) with the red dots ($\Nsp = 1024$) \footnote{Due to the high computational cost with more dust species, we have picked a smaller $T_{s,cut}$ sample with $\Nsp = 1024$ and hence fewer red points.}, it can be seen that the growth rate remains similar between the two cases when the lower cutoff of the distribution is high ($T_\mathrm{s,cut} \gtrsim 10^{-3}$). 
When $T_\mathrm{s,cut} \lesssim 10^{-3}$, the left panels show smaller growth rates with $\Nsp = 1024$ than with $\Nsp = 512$, indicating non-convergence, while the middle and the right panels continue to show consistent growth rates between $\Nsp = 512$ and $\Nsp = 1024$, indicating convergence.
Therefore, the two distinct regimes of convergence and non-convergence shown by Fig.~\ref{fig:multi1} are only manifested when sufficiently small dust species are included.

Moreover, it is not apparent that the largest particles alone can determine the maximum growth rate of multiple dust species in a distribution. The smaller particles contribute to the instability in a sophisticated way.
As shown in Fig.~\ref{fig:increasespe1}, the growth rate first increases when the next largest particles are activated in the distribution in both convergent and non-convergent cases. This can be understood as particles near $\sim \Tsmax$ contribute constructively to the instability. 
This trend continues until $T_\mathrm{s,cut} \simeq 0.7\Tsmax$ at which the growth rate reaches maximum except the case of $\varepsilon = 1$, $\Tsmax = 0.01$, and $q = -1.5$ whose rate reaches maximum at $T_\mathrm{s,cut}  \simeq 0.3\Tsmax$.
When even smaller particles are included, however, they can either constructively or destructively contribute to the instability.
We note also that the radial wave number often begins near the resonant drag regime $K_x \sim K_{a0}$ and shifts to the region with larger $K_x$ when sufficiently small dust particles are included (see Sec.~\ref{S:wavenumber}).

\begin{figure}
\includegraphics[trim=0mm 0mm 0mm 0mm, width=3.2in]{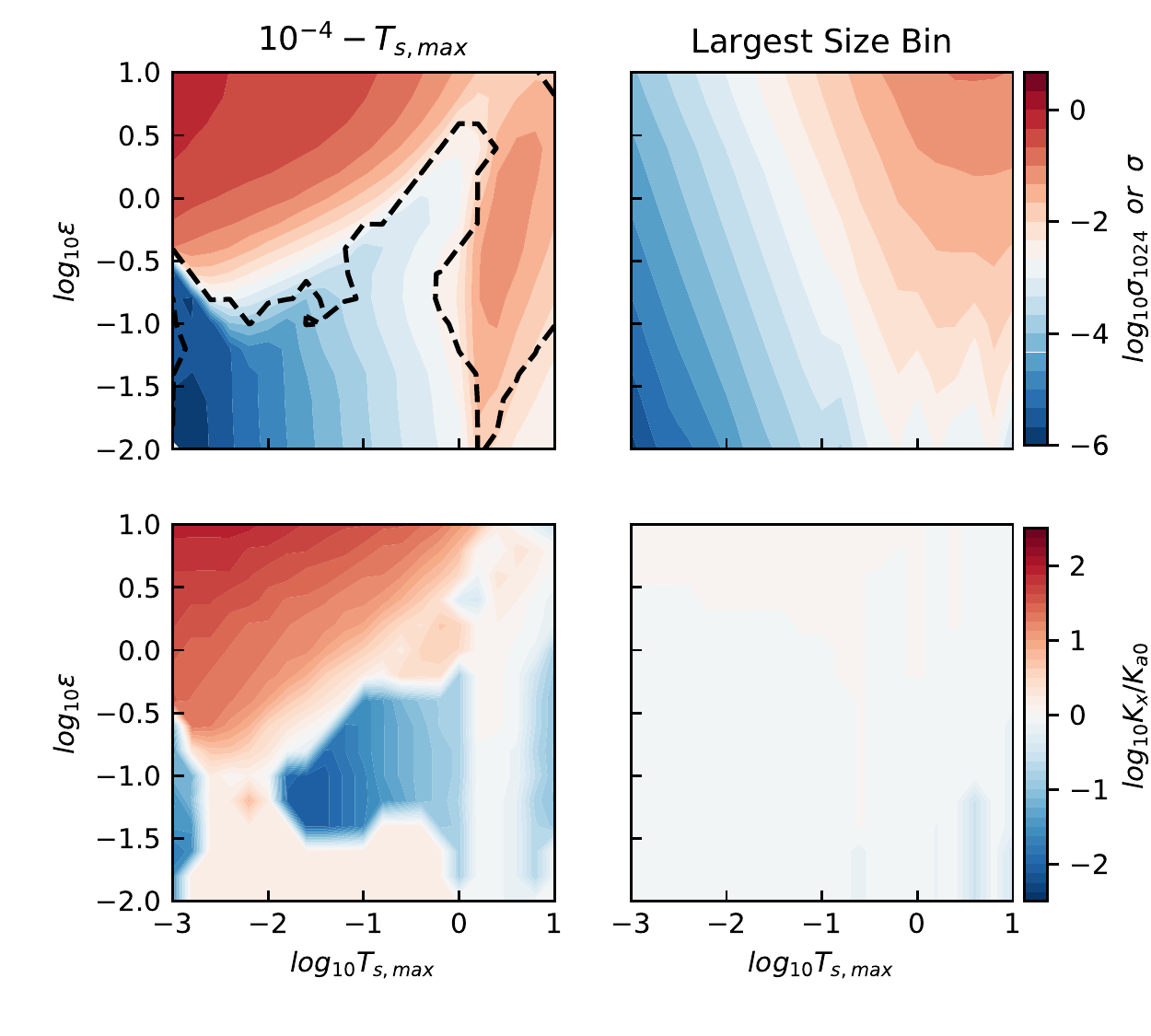}
\caption{%
 {Left panels: Similar to the $q = -3.5$ case in Fig.~\ref{fig:multi1} but derived by searching a wider $K_x$--$K_z$ domain (Sec.~\ref{SS:long}). 
Right panels: The maximum growth rate $\sigma$ and the dimensionless radial wave number $K_x$ of the most unstable mode for the single species whose $\varepsilon$ and $\Ts$ are from the largest size bin in the dust distribution used in the left panels. \label{fig:q3p5single}}}
\end{figure}

Overall, it is apparent that the small dust particles (e.g. $T_s\lesssim10^{-3}$) in a given distribution can have a strong effect on the instability, either positively or negatively.
The combined mass for dust with stopping times between $T_a$ and $T_b$ is
\begin{equation}
\frac{m(T_a<T<T_b)}{m_{total}}=\frac{T_b^{4+q}-T_a^{4+q}}{\Tsmax^{4+q}-10^{-4(4+q)}}.
\end{equation}
Therefore, dust with $\Ts \le 10^{-3}$  {constitutes} 24\% of the total dust mass for $\Tsmax = 0.01$ and $q = -3.5$ (upper panels) or only 0.3\% of the total dust mass for $\Tsmax = 0.01$ and $q = -1.5$ (lower panels). 
 {It can be seen} that, in the top middle panel, the growth rate jumps significantly at $T_\mathrm{s,cut} \simeq 10^{-3}$, which accompanies  {a} sudden change of $K_x$. This indicates that the most unstable mode suddenly jumps from one $K_x$-$K_z$ region to another region, and  {the contributions from the smaller particles may not simply be proportional}. 
In this regard, it is not necessary that the instability can be considered as if being driven by the largest dust particles, or that the smaller dust particles can only contribute to the instability negatively, as suggested by \cite{Krapp2019}.

 {To further demonstrate the differences between the streaming instability of multiple dust species and of one dust species representing the largest particle bin in a distribution alone, we consider the same $\Tsmax$--$\varepsilon$ space for our dust distributions as in Sec.~\ref{S:tworeg} but find the fastest growing mode for the latter scenario.
These two scenarios are compared in Fig.~\ref{fig:q3p5single}. We have adopted $\Nsp = 1024$ in the left panels.
Although the growth rates are both low at the small $\Ts$ and $\varepsilon$ corner and both high at the large $\Ts$ and $\varepsilon$ corner, they differ at low $\Ts$ and high $\varepsilon$.
More importantly, the fastest growing mode with single species is located at $K_x = K_{a,0}$ as expected (lower-right panel), but is significantly different from the one with multiple species (lower-left panel).
Therefore, the leading dust species does not appear to be acting independently in most of the parameter space we have considered.}

\subsection{Long-wavelength modes with \protect{$\Tsmax \lesssim 1$} and \protect{$\varepsilon \lesssim 1$}} \label{SS:long}
 {We note that for some cases, the $K_x$--$K_z$ domain we use to search for the fastest growing modes may be too restrictive.
This occurs near $\Tsmax \lesssim 1$ and $\varepsilon \lesssim 1$, i.e., around the blue region in the bottom panels of Fig.~\ref{fig:multi1}.
Figure~\ref{fig:test} suggests that the fastest growing modes are potentially located at even smaller $K_x$ and $K_z$ than we have considered.
Therefore, we repeat some of the previous calculations in this section with a larger search domain that extends into two orders of magnitude smaller wave numbers (see Sec.~\ref{SS:la}).}

 {The results for dust distributions with $\Tsmin = 10^{-4}$ but with different $q=-3.5$ and $q=-1.5$ are shown in the leftmost panels of Fig.~\ref{fig:q3p5multi} and~\ref{fig:q1p5multi}, respectively.
By comparing with the top-left and the top-right panels in Fig.~\ref{fig:multi1}, it can be seen that the fast-growth and the slow-growth regimes remain similar.
However, comparing the middle panels indicates that the region with convergent maximum growth rate is enlarged near $\Tsmax \lesssim 1$ and $\varepsilon \lesssim 1$.
This difference is due to the capture of the much smaller wave numbers, as shown by comparing the bottom panels.
Nevertheless, these long-wavelength modes remain slowly growing, may not fit into the thickness of a vertically stratified disc, and hence are not quite relevant in the context of protoplanetary discs (cf.\ Fig.~\ref{fig:multi2}).}

\subsection{Narrow dust-size distributions}\label{S:para}

\begin{figure*}
\includegraphics[trim=0mm 0mm 0mm 0mm, clip, width=6.6in]{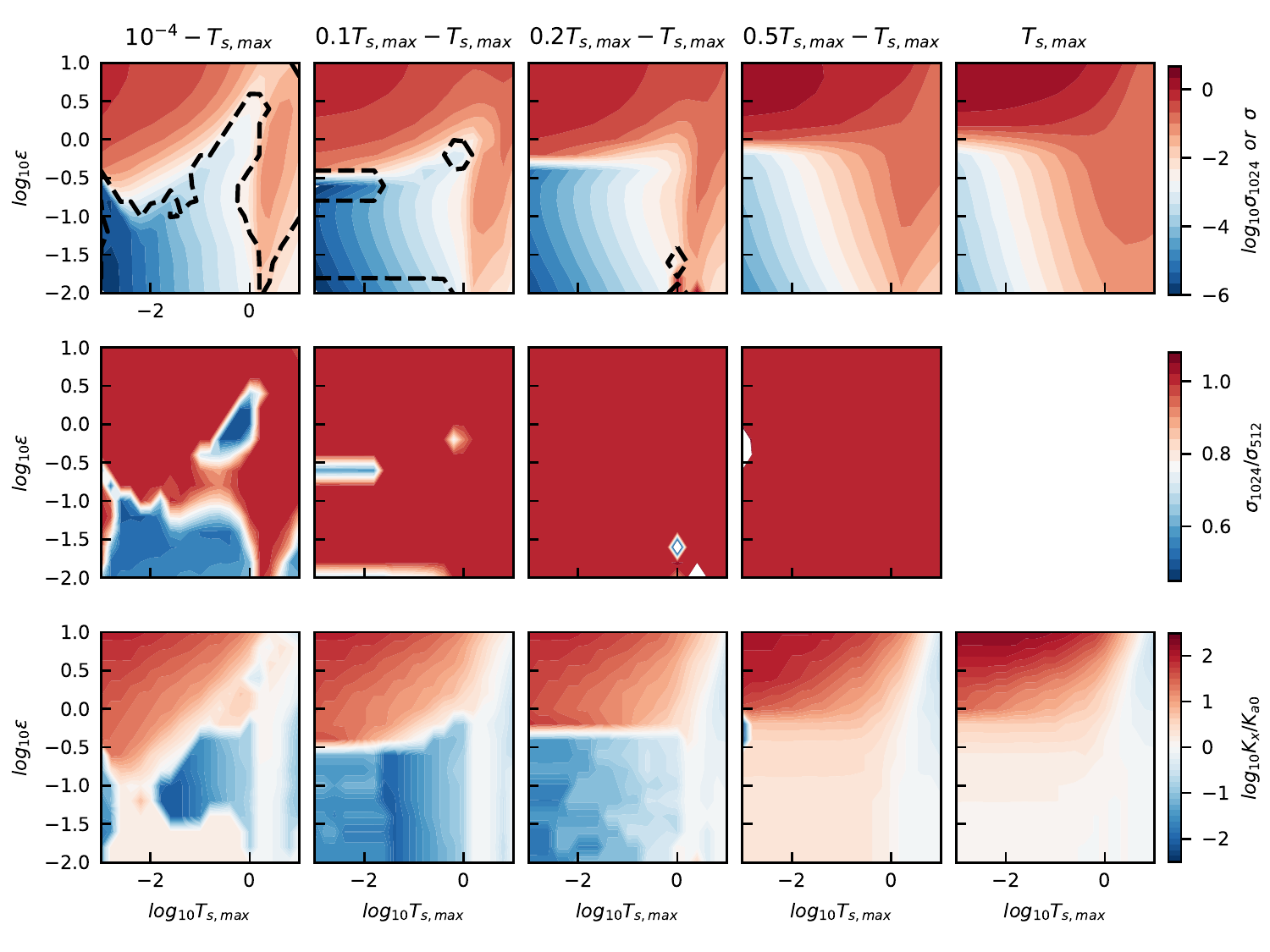}
\caption{%
  {%
  Similar to Fig.~\ref{fig:multi1} but with different $\Tsmin$.
 All dust distributions have $q=-3.5$. The rightmost panels show the most unstable modes with the single dust species at $\Ts = \Tsmax$.
 All the calculations use an extended $K_x$--$K_z$ search space for the fastest growing mode (Sec.~\ref{SS:long}).}
   \label{fig:q3p5multi}}
\end{figure*}

\begin{figure*}
\includegraphics[trim=0mm 0mm 0mm 0mm, clip, width=6.6in]{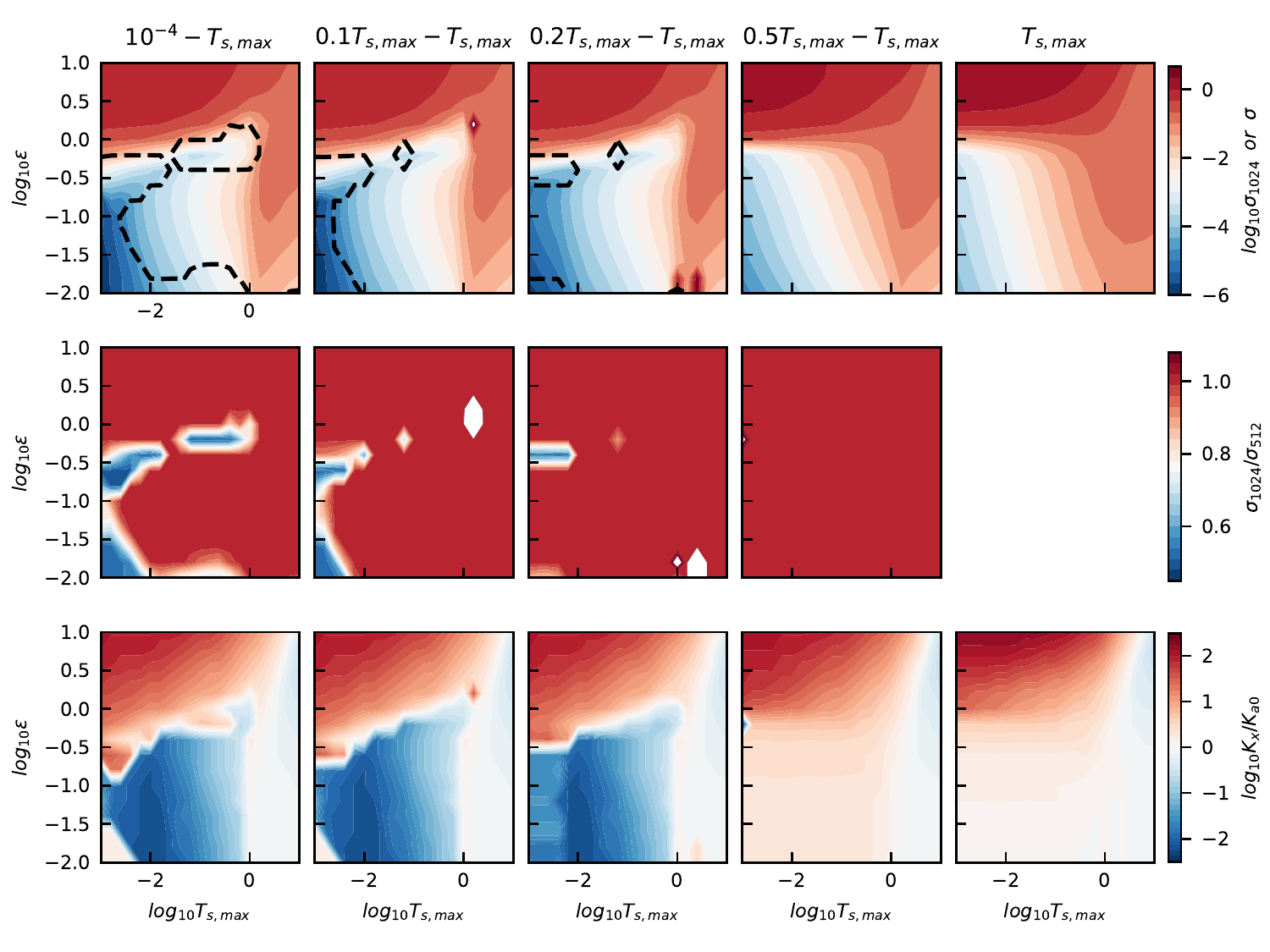}
\caption{%
  {%
 Similar to Fig.~\ref{fig:q3p5multi} but for dust distributions with $q=-1.5$.}
 \label{fig:q1p5multi}}
\end{figure*}

 {All our previous calculations assume that $\Tsmax$ and $\Tsmin$ ($\Tsmin=10^{-3}$ or $10^{-4}$) are well separated.
However, one could argue that dust grows preferentially to some characteristic sizes that are limited by growth barriers, so that $\Tsmax$ and $\Tsmin$ may be close to each other \citep[e.g.,][]{BDH08,Rozner2020}.
Furthermore, bringing $\Tsmin$ closer to $\Tsmax$ helps to shed some light on how the streaming instability with multiple species may make the transition into being considered as one with single species.
In this section, therefore, we vary $\Tsmin$ as a fraction of $\Tsmax$ and the results are shown in Figs.~\ref{fig:q3p5multi} and~\ref{fig:q1p5multi} for two different power-law index $q$.}

 {We find that a dust-size distribution need to be particularly narrow before it can be considered similar to a system with single species.
As shown by the top panels in Figs.~\ref{fig:q3p5multi} and~\ref{fig:q1p5multi}, the sharp boundary between the fast and the slow growth regimes ($\Tsmax\sim1$ and $\varepsilon\sim1$) remains apparent with $\Tsmin$ up to about $0.2\Tsmax$. 
This implies that, even with a dust-size distribution as narrow as $\sim$0.7\,dex, the dust particles do not interact coherently for the instability, and the unstable modes remain significantly different from those driven by single species.
Only when $\Tsmin \gtrsim 0.5\Tsmax$, it can be seen that all the dust act coherently and the growth rate becomes similar to that from single species. Note that all these calculates have adopted 512 and 1024 dust species except the rightmost panels.}

 {Another interesting trend is that the growth rate converges with the number of dust species $\Nsp$ in almost the entire $\Tsmax$--$\varepsilon$ domain when $\Tsmin\gtrsim0.1\Tsmax$ (see the middle panels of Figs.~\ref{fig:q3p5multi} and~\ref{fig:q1p5multi}).
Thus, unlike previous cases of wide dust-size range where the fast and slow growth rates are in the converged and non-converged regimes separately, both fast and slow growth rates are converged with the number of dust species when $\Tsmin$ is close to $\Tsmax$.}

 {These results are consistent with Fig.~\ref{fig:increasespe1} (discussed in Sec.~\ref{S:interaction}) where adding particles with  $\Ts \gtrsim 0.5\Tsmax$ act coherently to increase the growth rate of the instability while adding even smaller particles starts to show deviation from the coherence.
Eventually, adding particles with $\Ts \lesssim 0.2\Tsmax$ can result in non-converged growth rates with number of dust species (the left panels in Fig.~\ref{fig:increasespe1}).}

\section{Summary} \label{S:summary}
We have conducted analytical calculations and direct numerical simulations to study the linear phase of the streaming instability with multiple dust species.
We have explored various dust distributions with different maximum dust sizes $\Tsmax$, total solid-to-gas mass ratios $\varepsilon$, and power-law indices $q$.
In general, the instability has a high growth rate when \emph{either} the maximum size $\Tsmax$ \emph{or} the total solid-to-gas mass ratio $\varepsilon$ is high ($\Tsmax\gtrsim1$ or $\varepsilon\gtrsim1$).
Under these conditions, the growth rate is converged with increasing number of dust species $\Nsp$ representing the distribution.
 {By contrast, the growth rate in the complimentary region of the parameter space, i.e., $\Tsmax\lesssim1$ \emph{and} $\varepsilon\lesssim1$, is quite low as long as $\Tsmin\lesssim0.2 \Tsmax$.
When $\Tsmin$ is $10^{-3}$ or $10^{-4}$, the growth rate continues to decrease significantly in this $\varepsilon$--$\Tsmax$ space with successively larger $\Nsp$, implying that the growth rate with the limit of a continuous dust size distribution can be even lower than our calculated rates.
When the size distribution is as narrow as $\Tsmin\gtrsim 0.1\Tsmax$, the growth rate seems to converge, albeit low.}
For more top-heavy dust distributions (e.g.\ $q = -1.5$), there is a larger parameter space for high growth rates extending down to $\Tsmax \sim 0.1$ (but the fastest growing modes for $0.1\lesssim\Tsmax\lesssim1$ cases may not fit in the thickness of a protoplanetary disc).
Overall, the growth rate is clearly separated into the fast and  {the} slow regimes when multiple dust species have been considered as long as $\Tsmin\lesssim0.2\Tsmax$.

We find that the transition between the converged (fast-growth) and the non-converged (slow-growth) regimes is excessively sharp when $\Tsmin=10^{-4}$.
The more dust species $\Nsp$, the sharper the transition becomes.
Therefore, the growth rate map with respect to $\Tsmax$ and $\varepsilon$ is clearly separated into converged and non-converged regions by a discontinuity-like boundary (Figs.~\ref{fig:multi1} and~\ref{fig:multimin1em3}).
Interestingly, for $\varepsilon\gtrsim 1$, the most unstable mode in the converged region has a dimensionless radial wave number $K_x \gtrsim 10 K_{a0}$, which is in a different branch from $K_x \sim K_{a0}$ identified as the resonant drag instability driven by the largest species \citep{SH18}.
For $\varepsilon\lesssim 1$ and $\Tsmax\gtrsim 0.1$, on the other hand, the most unstable mode in the converged region has a variety of $K_x$ depending on $\Tsmax$.
When $\Tsmax\gtrsim1$, $K_x$ is close to $K_{a0}$.
When $\Tsmax$ is small ($0.1\lesssim\Tsmax\lesssim1$), the fastest growing mode has $K_x\ll K_{a0}$, but these modes also have small $K_z$ such that they cannot fit into typical disc thickness and should not operate in protoplanetary discs.

We also notice that, for a disc with a given $\varepsilon$ and $\Tsmax$, only unstable modes in a limited range of wave numbers have converged non-zero growth rates with increasing number of dust species $\Nsp$.
Thus, when the disc condition changes (either $\varepsilon$ or $\Tsmax$), the most unstable mode can move from outside the converged $K_x$--$K_z$ region into the converged region.
In this case, the growth rate of the instability changes from being non-convergent to convergent.
This may help explain the sharp transition between the converged and the non-converged regions in the $\varepsilon$--$\Tsmax$ growth rate maps.  

Different dust species in a distribution appear to interact and contribute to the instability in non-trivial ways.
We find that for a wide range of conditions, the dynamics may not be simplified to the one driven by the largest species in isolation, as may have been suggested by \cite{BS10}, \cite{SYJ18}, and \cite{Krapp2019}.
Smaller dust species can either contribute positively or negatively to the instability.
Moreover, they can change the relative importance of the unstable modes between different branches in the Fourier space to which the growth rate depends on sensitively (Figs.~\ref{fig:test} and~\ref{fig:test2}).


Finally, we have used hybrid numerical simulations to reproduce the unstable modes for several representative cases.
Numerical convergence has been achieved in all cases, and the corresponding analytical growth rates down to unprecedentedly low $\sigma \simeq 10^{-3}\OmegaK$ as well as for large leading species ($\Tsmax \sim 2$) are recovered.
In the process, we have demonstrated the resolution requirement for simulating the instability with the code, which generally lies within 16--64 grid points per wavelength of the unstable modes.
This experiment will serve us for future investigations of the streaming instability with multiple dust species using direct numerical simulations.

\section*{Acknowledgments}
All simulations are carried out using computer supported by the 
Texas Advanced Computing Center (TACC) 
at The University of Texas at Austin through XSEDE grant TG-AST130002 and from the NASA High-End Computing (HEC) Program through the NASA Advanced Supercomputing (NAS) Division at Ames Research Center. 
ZZ acknowledges support from the National Science Foundation under CAREER Grant Number AST-1753168 and NASA
TCAN award 80NSSC19K0639.
CCY is grateful for the support from NASA via the Emerging Worlds program (Grant \#80NSSC20K0347).
We thank Leonardo Krapp, Andrew Youdin, and other TCAN members for constructive comments during the 2019 Tucson TCAN meeting and subsequent discussions/comments.

\section*{DATA AVAILABILITY}
The data underlying this article are available in the article and in its online supplementary material.

\bibliographystyle{mnras}
\input{msresubmit.bbl}

\appendix

\onecolumn
\section{Seeding density perturbations by Lagrangian particles} \label{S:seedpar}

\cite{YJ07} in their Appendix~C described how to seed a sinusoidal perturbation in density with Lagrangian particles.
When there exist multiple perturbations with an arbitrary phase, cross correction terms are needed to maintain high order of accuracy.
In this appendix, we illustrate how this can be achieved.

Following \cite{YJ07}, we consider the superposition of two waves with wave numbers $\vec{k}_\pm \equiv k_{x,0}\unitvec{x} \pm k_{z,0}\unitvec{z}$, resulting in one horizontally propagating but vertically standing wave.
In this case, the initial density field of the particles in the $xz$-plane reads (see their Equation~(10a))
\begin{align}
\rhop(\vec{r})
=\,& \rho_0\left[1 + \left(\AR \cos k_{x,0} x - \AI\sin k_{x,0} x\right)\cos k_{z,0}
         \right]\label{E:rhop1}\\
=\,& \rho_0\left[1 + \frac{1}{2}\AR
                     \left(\cos\kp\cdot\vec{r} + \cos\km\cdot\vec{r}\right)
                   - \frac{1}{2}\AI
                     \left(\sin\kp\cdot\vec{r} + \sin\km\cdot\vec{r}\right)
           \right],\label{E:rhop2}
\end{align}
where $\vec{r} \equiv x\unitvec{x} + z\unitvec{z}$ is the position, $\rho_0$ is the mean density, $A \equiv \AR + i\AI$ is the complex amplitude of the waves \emph{relative} to $\rho_0$.

To approximate the continuous density field (Equation~\eqref{E:rhop1}) using discrete particles, we begin with a periodic domain accommodating integral numbers of wavelengths $\lambda_{x,0} \equiv 2\pi / k_{x,0}$ and $\lambda_{z,0} \equiv 2\pi / k_{z,0}$ in both directions.
We uniformly distribute $N_\mathrm{p}$ identical particles with positions $\rj$ and shift the position of each particle by $\xj$.
The density field of the particles is then
\begin{equation} \label{E:disc_rhop}
\rhop(\vec{r}) = \mass\sum_j\delta\left(\vec{r} - \vec{r}_j - \xj\right),
\end{equation}
where $\mass$ is the mass of each particle and $\delta(\vec{r})$ is the Dirac $\delta$ function.%
\footnote{We note that any of the weighting schemes of the particle-mesh method has the properties of a $\delta$ function.}
Fourier transforming Equation~\eqref{E:disc_rhop} over the domain gives
\begin{equation} \label{E:rhopk}
\widetilde{\rhop}(\vec{k})
= \mass\sum_j \exp\left(i\vec{k}\cdot\vec{r}_j\right)
              \left[1 + i\vec{k}\cdot\xj
                      - \frac{1}{2}\left(\vec{k}\cdot\xj\right)^2
                      + O\left(\left|\vec{k}\cdot\xj\right|^3\right)\right].
\end{equation}

Equation~\eqref{E:rhop2} suggests that to first order in $|A|$, the position shifts should be
\begin{equation}
\xj^{(1)} =
-\frac{\AR}{2k_0^2}\left(\kp\sin\kp\cdot\rj + \km\sin\km\cdot\rj\right)
-\frac{\AI}{2k_0^2}\left(\kp\cos\kp\cdot\rj + \km\cos\km\cdot\rj\right),
\end{equation}
where $k_0 \equiv \sqrt{k_{x,0}^2 + k_{y,0}^2}$.
Substituting $\xj$ = $\xj^{(1)}$ into Equation~\eqref{E:rhopk} results in
\begin{align}
\widetilde{\rhop}(\vec{k}) =\,&
N_\mathrm{p}\mass\left[1
    +\frac{1}{4}\AR\left(\delta_{\vec{k},+\kp} + \delta_{\vec{k},-\kp}
                       + \delta_{\vec{k},+\km} + \delta_{\vec{k},-\km}
                   \right)
    -\frac{i}{4}\AI\left(\delta_{\vec{k},+\kp} - \delta_{\vec{k},-\kp}
                       + \delta_{\vec{k},+\km} - \delta_{\vec{k},-\km}
                   \right)\right.\nonumber\\
    &+\frac{1}{8}\left(\AR^2 - \AI^2\right)
                \left(\delta_{\vec{k},+2\kp} + \delta_{\vec{k},-2\kp}
                    + \delta_{\vec{k},+2\km} + \delta_{\vec{k},-2\km}
                \right)
    -\frac{i}{4}\AR\AI\left(\delta_{\vec{k},+2\kp} - \delta_{\vec{k},-2\kp}
                            + \delta_{\vec{k},+2\km} - \delta_{\vec{k},-2\km}
                        \right)\nonumber\\
    &+\frac{1}{4}\left(\AR^2 + \AI^2\right)\left(\frac{k_{z,0}}{k_0}\right)^4
                  \left(\delta_{\vec{k},+2k_{z,0}\unitvec{z}}
                      + \delta_{\vec{k},-2k_{z,0}\unitvec{z}}\right)
    +\frac{1}{4}\left(\AR^2 - \AI^2\right)\left(\frac{k_{x,0}}{k_0}\right)^4
                  \left(\delta_{\vec{k},+2k_{x,0}\unitvec{x}}
                      + \delta_{\vec{k},-2k_{x,0}\unitvec{x}}\right)\nonumber\\
    &\left.
      -\frac{i}{2}\AR\AI\left(\frac{k_{x,0}}{k_0}\right)^4
                  \left(\delta_{\vec{k},+2k_{x,0}\unitvec{x}}
                      - \delta_{\vec{k},-2k_{x,0}\unitvec{x}}\right)
    + O(|A|^3)\right],
\end{align}
where $\delta_{\vec{u},\vec{v}} \equiv \delta(\vec{u} - \vec{v})$ for any $\vec{u}$ and $\vec{v}$.
In other words, perturbations of amplitude on the order of $|A|^2$ appear in the second harmonics.
These overtones can be cancelled by introducing the second-order correction shifts
\begin{align}
\xj^{(2)}
=\,& \frac{\AR^2 - \AI^2}{8k_0^2}\left(\kp\sin2\kp\cdot\rj
                                     + \km\sin2\km\cdot\rj\right)
    +\frac{\AR\AI}{4k_0^2}\left(\kp\cos2\kp\cdot\rj
                                     + \km\cos2\km\cdot\rj\right)\nonumber\\
   &+\frac{\left(\AR^2 + \AI^2\right)k_{z,0}^3}{4k_0^4}\unitvec{z}\sin2k_{z,0}z_j
    +\frac{\left(\AR^2 - \AI^2\right)k_{x,0}^3}{4k_0^4}\unitvec{x}\sin2k_{x,0}x_j
    +\frac{\AR\AI k_{x,0}^3}{2k_0^4}\unitvec{x}\cos2k_{x,0}x_j.
\end{align}

\bsp
\label{lastpage}
\end{document}